\documentclass[reprint,longbibliography,pre,amsmath,amssymb,amsfont,url, aps]{revtex4-2}
\usepackage{bold-extra}
\usepackage{amsthm}
\usepackage{dutchcal}

\usepackage[colorlinks, allcolors=blue]{hyperref}
\usepackage{graphicx} 
\usepackage{csquotes}
\usepackage{mathtools}

\usepackage{enumitem}
\setlist[enumerate]{parsep=3pt}

\usepackage{booktabs}

\newcommand{\abs}[1]{\lvert #1 \rvert}
\newcommand{\myquote}[1]{\textit{\textquote{#1}}}
\newcommand*\diff{\mathop{}\!\mathrm{d}}
\newcommand*\vx{\textbf x}
\newcommand*\W{\mathcal W}

\newcommand{\law}[2]{
	\vspace{0.25cm}
	\noindent
	\textbf{#1}: 
	\textit{#2}
	\vspace{0.2cm}
}

\begin{document}
	\author{D. Lairez}
	\email{didier.lairez@polytechnique.edu}
	\affiliation{Laboratoire des solides irradi\'es, \'Ecole polytechnique, CEA, CNRS, IPP,
		91128 Palaiseau, France}
	\title{Energy and information:\\ a chronicle of hesitations on the role of the observer in physics}
	\date{\today}
	
	\begin{abstract}

Energy has no definition, except that given by a conservation principle which essentially amounts to defining it as the elements of an open list of unknown cardinality. Entropy, identified by Shannon as information we lack, has too many definitions. This results in an unstable and hesitant interpretation of their link.

Thermodynamics, the science of changes in form of energy, is phenomenological, all its laws are induced from observation. From the origin, the concept of energy is linked to the observer's knowledge, to the information he has: what and where to look and with what instruments. 
Thermodynamics only addresses the sensible world. It is Aristotelian. 
But this is disturbing if we consider that reason can give us access to Plato's intelligible world, the one that is beyond the sensible world and independent of us. This is disturbing if we consider that science can access to the intrinsic properties of things, those which are independent of us.
This is disturbing if we have a purely Platonic conception of science. Hence the statistical mechanics approach ("The rational foundation of thermodynamics", J.W. Gibbs).
This is the first pendulum movement of ideas, whose oscillations continue to this day, because unfortunately statistical mechanics introduces many inconsistencies, mainly due to the ergodic hypothesis. Luckily, these inconsistencies are all solved by Shannon's information theory. Sadly, information theory is too Aristotelian and too conceptual. Fortunately, Landauer principle makes it more \textquote{physical}. This is currently the latest attempt to bringing the notions of energy and information back to what is considered the right side of science, that of Plato. Landauer principle is now commonly regarded as a fundamental law of physics. Unpleasantly, it can be shown that this principle is not one.
			
	\end{abstract}
	
	\maketitle
	
	\section*{Introduction}
	
The link between energy and information is \textit{a priori} confusing because of the great influence of Descartes philosophy on modern science and the object/subject duality (or body/mind duality) that he explicitly introduced\,\cite{Descartes2008, Carter_1931}. On the one hand, energy is one of the most fundamental concepts in physics and the physical world is usually viewed as an object, that is to say something that is observed. On the other hand, information is “the imparting of knowledge in general” (Oxford Dictionary) and knowledge refers to the human as a subject, that is to say a being that
is thinking and observing, in brief an observer.
The link observed/observer is clear but it becomes disturbing when we say that information is energy. There is confusion of roles.

In a quite general manner, a scientific theory is usually viewed as a consistent set of concepts accounting for objective reality (that which we can know)\,\cite{Mach_1919, Duhem_2021, Einstein_1934, Einstein1935}. Everything should be clear. Except that science does not deal with the essence of thing, so that the most fundamental concepts has no definition (this is the case of energy). And except that science also does not deal with the existence of thing, so what reality means is actually a matter of personal viewpoint that has been a subject of debate from ancient Greece to the present day.

Broadly speaking, the different points of view fall into two categories, which can be symbolized by that of Plato and that of Aristotle:
\begin{itemize}
	\item Plato: we can know through reason what exists independently of humans. This \textit{intelligible world}, which exists without any observer, is the objective reality with which science deals.
	\item Aristotle: what we can know is only the part of the world with which we interact, that is, the part that is observable. This \textit{sensible world} common to all is the objective reality with which science deals.
\end{itemize}
The object/subject duality has its roots in this very old debate.
The Plato's \textit{intelligible world} is objective by definition. The way in which the \textit{sensible world} becomes objective is less direct. In this latter case, \textquote{objective} means \textquote{non-personal}, that is, \textquote{independent of the person of the observer}. In other words, an observation must be reproducible by anyone with the same means and the same information.
The objectivization of the \textit{sensible world} necessarily involves the concept of information.

Both points of view are defensible and this article will certainly not close the debate. Nowadays mathematics (and so logic) are usually viewed as purely Platonic, whereas natural sciences are usually perceived as Aristotelian because observations and experiments are the ultimate judges of truth: \myquote{all knowledge about reality begins with experience and terminates in it} (A. Einstein\,\cite{Einstein_1934}). 
However, even the natural sciences use mathematics and are submitted to the imperative of logical consistency and embed a Platonic element.
 Thus, logical-empiricism, or neo-positivism, or logical-positivism, which recognizes the two contributions: logic and observations as the only sources of knowledge, should be, it seems to me, fairly consensual. In any case, whatever one's personal opinion on this debate, classical logic, and more particularly its law of non-contradiction, being the only and ultimate common point on which everyone agrees, provided that this logic is preserved, neither option is more scientific than the other.

\begin{figure}[!htbp]
	\begin{center}
		\includegraphics[width=1\linewidth]{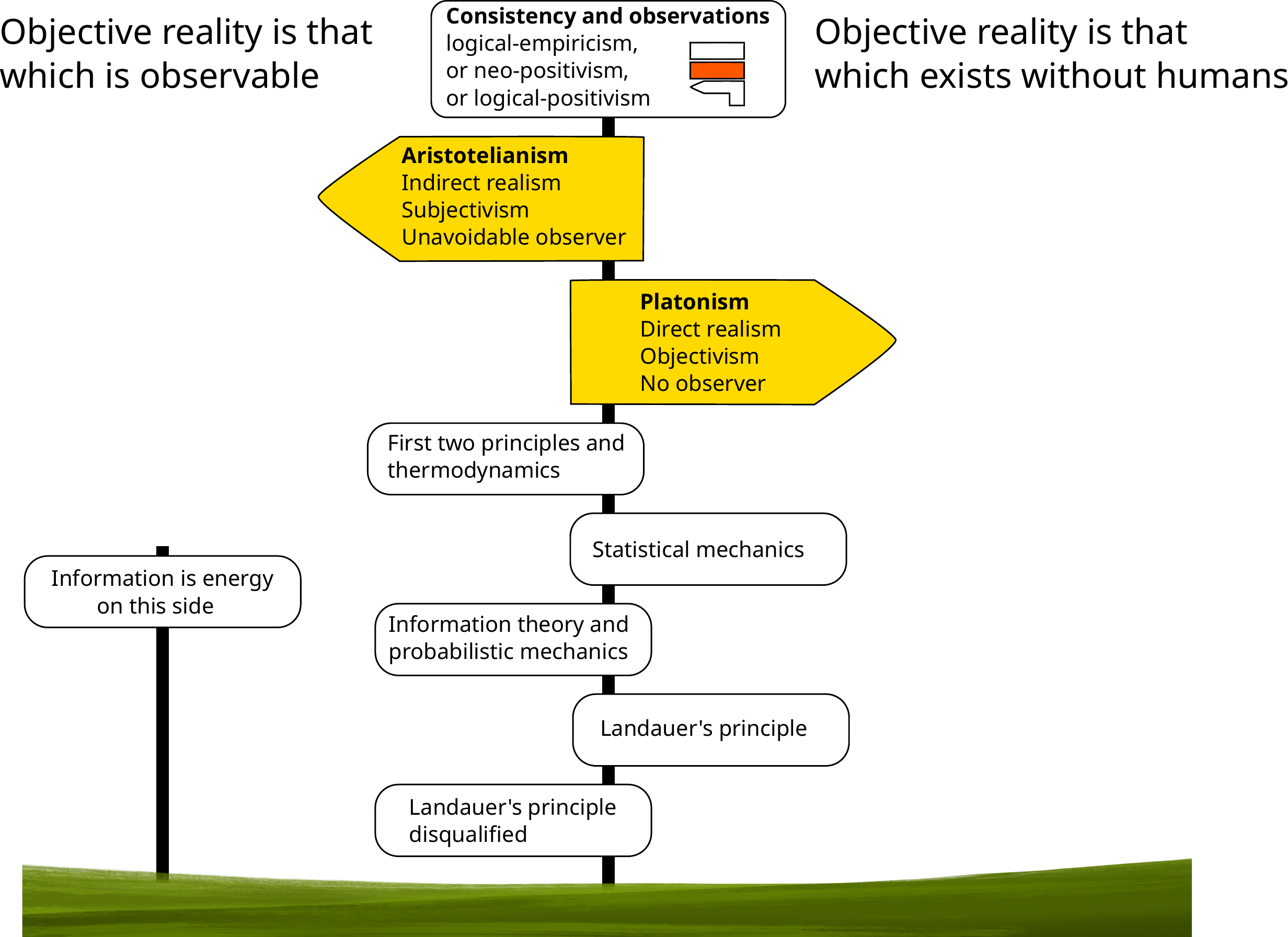}
		\caption{Road-map of the paper: submitted to two alternating and opposite forces, the position of the observer has oscillated from the origin of the concept of energy to the present day.}
		\label{panneau}
	\end{center}
\end{figure}

The aim of this paper is to show how, despite what we have just said, on the question of energy and thermodynamics the role of the observer and his knowledge is regularly denied with the only implicit argument being that of a pure Platonism. This comes at the cost of many inconsistencies that are regularly resolved by reintroducing the observer into the problem again.
Submitted to these two alternating forces, the position of the observer has oscillated from the origin of the concept of energy to the present day (Fig.\,\ref{panneau}).
The purpose of this article is to retrace this chronicle.
It is divided in five sections that follow the chronology: thermodynamics and its first two principles, statistical mechanics, Shannon's information theory, Landauer principle and its derivatives and finally the invalidation of the latter.

\tableofcontents

\newpage

\section{Thermodynamics and the first two principles}

The concept of energy comes from thermodynamics, which is probably the most emblematic example of what a phenomenological theory is. Everything starts from observations and experiments to which a consistent set of general laws is derived by induction\,\cite{Maxwell_1872, Planck_1903}.
But thermodynamics is and remains phenomenological. That is to say, the concepts used in thermodynamics are highly dependent on observations and on the prior information needed for those observations: what and where to look and with what instruments. 

The famous quote \myquote{The idea of dissipation of energy depends on the extent of our knowledge} (J. C. Maxwell\,\cite{Maxwell_1878}), clearly refers to the second principle and entropy (dissipation) and 
puts the observer at a privileged position.
But actually this position was already established from the first principle. So that, thermodynamics as a whole distances physics from a purely Platonic conception of \textit{objective reality}.

\subsection{First principle: energy, changes and equilibrium}\label{first_principle}

Before Joule, we knew about heat, which is what is necessary to raise the temperature of a body, and we knew about mechanical work, which is what is necessary to set it in motion. But anyone who has ever warmed their hands by rubbing them together knows that mechanical work can turn into heat. Joule\,\cite{Joule_1850} showed that the quantity of heat so produced is always proportional to the quantity of mechanical work provided. So that by choosing a correct unit for both, these two quantities are equal. A new physical quantity was born: energy. It was not discovered, but invented.

However, \myquote{it is important to realize that in physics today, we have no knowledge of what energy is.} (R. Feynmann\,\cite{Feynmann_Energy}).
Energy is a pure abstraction. There is no such thing as an \textquote{energy particle} (an object) that could exist independently of us and independently of our thinking. If energy is considered as belonging to \textit{the objective reality}, this reality is not that of Plato.

Energy is only \textquote{defined} by a principle of conservation: the first principle. But this is not a definition that would allow us, for example, to recognize energy with certainty when we see it. In classical mechanics, we have kinetic energy and potential energy, the latter being \textquote{potential} in the sense that it is potentially kinetic energy. In special relativity, we have energy of motion (kinetic energy) and energy of rest (rest mass), the latter being here again defined by opposition to the former. In thermodynamics, energy is transferred under the form of heat or under the form of work (anything that is not heat), and energy is stored under the form of thermal energy (identified later as corresponding to the average kinetic energy of particles) or under the form of anything that is not thermal energy.
Clearly, the only thing we can recognize with certainty is kinetic energy. The other forms of energy are \textquote{defined} as not being kinetic energy, they are defined as a sort of complementary part of the latter. But it is a complementary part based on a whole that is not closed and always opened to the discovery of new forms of energy. 
In case we discover that a phenomenon produces a non-zero energy balance, it means that a new form of energy has been discovered. Since the birth of this concept, this is how it has always worked and there is actually no alternative that would be logically consistent.
This is a key point. 
The principle of conservation is inseparable from leaving open the list of what is considered to be energy.
Closing this list would amount to consider as impossible new discoveries. But by definition, we do not know the future of our knowledge, so that new discoveries cannot be excluded. Hence the inconsistency that would consist is closing the list.

\begin{figure}[!htbp]
	\begin{center}
		\includegraphics[width=0.4\linewidth]{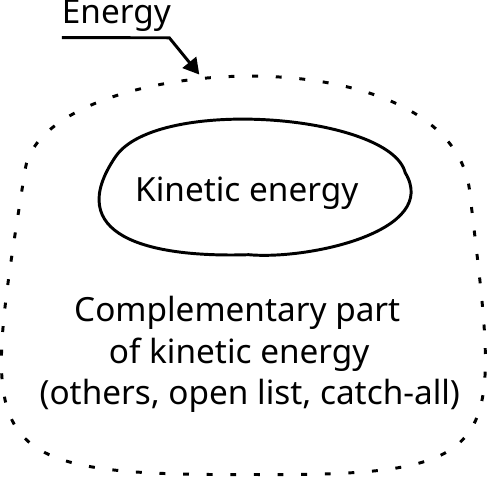}
		\caption{Principle of conservation: energy is the thing that is conserved when everything else change. If a phenomenon produces a non-zero energy balance, it means that a new form of energy has been discovered. We known kinetic energy. the others form a list that remains open to new discovery.}
		\label{energy}
	\end{center}
\end{figure}

In the absence of anything better, energy is \textquote{defined} as the thing that is conserved. But like that, this operational definition is incomplete, because the idea of conservation makes sense in relation to change.
So that, energy is the thing that is conserved when everything else change.
This is in line with Planck's idea of defining the energy of a body \myquote{as the faculty to produce external effects} \,\cite{Planck_1903}-\S56, as \textquote{effects} are only evaluated by changes (for a very interesting paper on the link between energy and changes see\,\cite{Hecht_2019b}).

The concept of energy is therefore in reality inseparable from that of change, which is itself inseparable from that of time. This brings us to the definition of equilibrium in thermodynamics which is conceived as the stationary state in which no change occurs.
An important point for the sequel deserves to be underlined. In phenomenological thermodynamics, there are no fluctuations in the theory, nothing spontaneous takes the system out of its equilibrium, which therefore does not need to be restored. Equilibrium does not need to be an attractor for the theory to be consistent. The equilibrium is the optimum of nothing at all.
It was only after the advent of statistical mechanics that this notion of optimum appeared.

\subsection{Second principle: asymmetry of changes, irreversibility and Clausius entropy}

While mechanical work can be completely changed into heat, the reverse is always incomplete\,\cite{Planck_1903}.
Relative to the energy involved, any change of a system in a given direction, say from state A to state B, is never the symmetrical of the reverse that brings the system from B to A. The system always provides more heat to the surroundings from A to B than the surroundings to the system from B to A. 
The system always supplies more heat to the environment on the way there, than the environment supplies to the system on the way back.
Regardless of the changes that occur during a cycle, part of energy is always dissipated as heat transferred to the surroundings\,:
\begin{equation}\label{Clausius2}
	-Q_{AA}\ge 0
\end{equation}
where the subscript $\text{AA}$ designates the cycle $\text{A}\to\text{B}\to\text{A}$, the sign of $Q_{AA}$ is relative to the system and $-Q_{AA}$ is the quantity of heat dissipated.
In Eq.\,\ref{Clausius2}, the equality sign holds for ideal reversible processes, which strictly speaking never exist in reality.
For real systems, the heat dissipated for a complete cycle is always strictly positive but tends towards zero as the evolution rate tends also towards zero. These quasi-static systems, although rigorously speaking irreversible, are called reversible for convenience of language. Because there exist free processes, that run out of control and which rate of evolution cannot be driven, so that the heat dissipated is always finite. These are the ones we call irreversible.

This inevitable heat dissipation is accounted for by the second law of thermodynamics that introduces a new state function $S$, named entropy\,\cite{Clausius_1879} (which will be referred to as Clausius entropy hereafter),
such as, whatever the change of state that a system undergoes at constant thermal energy $T$, we always have:
\begin{equation}\label{Clausius_ineq1}
	-Q \ge -T\Delta S
\end{equation}
where $-Q$ is the quantity of heat given by the system and received by the surroundings. Eq.\,\ref{Clausius_ineq1} is the Clausius inequality. 

The equality sign in Eq.\,\ref{Clausius_ineq1} holds for a reversible ideal change of state (whose path expressed by a differential form can be followed infinitely slowly in a quasistatic manner) and serves as a definition of entropy in thermodynamics:
\begin{equation}\label{Clausius_definition}
	\diff S = \frac{\diff Q}{T}
\end{equation}

The inequality sign in Eq.\,\ref{Clausius_ineq1} is the general case.
At constant internal energy, the sum of heat $Q$ and work $W$ received by the system is zero: $Q+W=0$, and Eq.\,\ref{Clausius_ineq1} can be rewritten as
\begin{equation}\label{Clausius_ineq2}
	W \ge -T\Delta S
\end{equation}
where $W$ is the work required to change the state of the system toward a state of lower entropy.
In other words (that will be useful in the sequel), any negative variation of entropy of the system requires a minimum work $-T\Delta S$ to be provided by the surroundings.


\subsection{The role of the observer} \label{observer_thermo}

Thermodynamics is phenomenological, so that by definition it places the observer in a key position. Thermodynamics deals only with phenomena. This means that thermodynamics deals only with observable changes, but also that in thermodynamics there is no change that is not observable. The \textit{objective reality} of thermodynamics is Aristotelian.

The consequence, which seems to be a truism, is that in thermodynamics, whether two states are different or not depends on the ability to distinguish changes and differences. For example, it depends on the resolution of our instruments, it depends of detection threshold of impurities etc. It depends also on our current knowledge, for example there was no change in isotopic composition before the discovery of isotopes\,\cite{Jaynes1992}.
In the case where two states are perceived as identical, no work is necessary to move the system from one to the other because there is only one.
Let us take two examples.

\subsubsection{Free expansion}\label{free_exp}

\begin{figure}[!htbp]
	\begin{center}
		\includegraphics[width=0.8\linewidth]{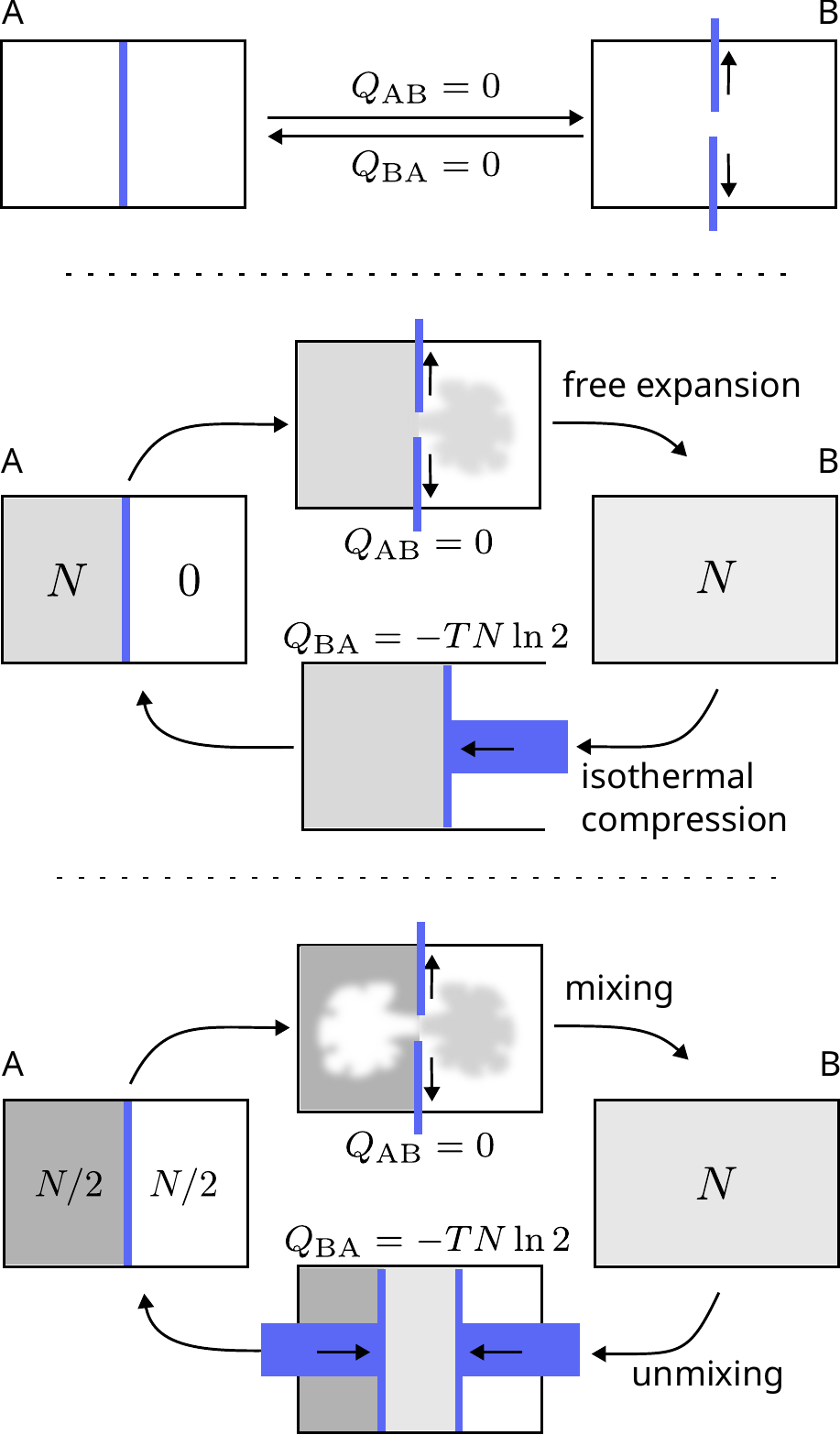}
		\caption{Top: joining two volumes of which we have no information on their contents: empty or filled with the same gas species; or one filled with a gas the other empty; or one filled with one type of gas the other with another type etc. The process is thermodynamically reversible. Replacing the partition leaves the system in the same state as before it was removed.\\
		Middle (free expansion): same operation, but this time we know that one volume is originally filled with $N$ particles of gas. The process is thermodynamically irreversible because it requires an isothermal compression with a piston in order to restore the initial state. So that the entire cycle dissipates $TN\ln2$ of heat.\\
		Bottom (mixing): same operation, but this time we know that the two compartments are originally filled with two different gas species. The process is thermodynamically irreversible because it requires the isothermal compression of each species (by the mean of specific semi-permeable membranes\,\cite{Planck_1903}).}
		\label{freeexp1}
	\end{center}
\end{figure}

Consider a container equipped with a wall that divides it in two and can be removed and replaced without friction (Fig.\,\ref{freeexp1}).
Let us call A the initial state with the partition. Remove the partition (state B), then replace it and compare this final state A' to state A. 
Any difference between states A and A' depends on what we know about these states and on what we are able to measure. Let us assume that we are able to measure the energy exchanges between the container and the surroundings (heat or work), but that in this case we did not measured anything\,:
\begin{enumerate}
	\item If the information we have is limited to what we have just said, then states A and A' are identical. No further operation or additional work is required to return to the initial state and a thermodynamic cycle is closed (Fig.\,\ref{freeexp1}, top).
	
	\item If we have the additional information that initially in state A, the left side of the container was filled with $N$ particles of gas, while the right side was empty (this assumes that we have at our disposal a detector capable of quantifying the amount of gas), then A and A' differ. An additional mechanical work is necessary to compress the gas by a factor $2$ so that it returns into its initial compartment and the system is restored to state A. At constant temperature, this work dissipates at least $TN\ln 2$ of heat into the surroundings (Fig.\,\ref{freeexp1}, middle).
\end{enumerate}

The first step of this example, that consists in removing the partition, is the  free expansion of a gas. It is not accompanied by any exchange of heat or work, the internal energy is constant and the thermal energy (the temperature) of the gas does not change. So, if these are the only sources of information (first case), nothing has happened and this first step of the cycle is entirely reversible, because we do not know that it was in fact a free expansion. If we have an additional source of information (second case with a gas detector), then  during this first step something has happened that requires additional work to be reversed. The first step (the free expansion) is then thermodynamically irreversible.

The thermodynamic irreversibility of a process depends on our knowledge.

\subsubsection{Gibbs paradox of mixing}\label{Gibbs_paradox1}

The free expansion of a volume of gas into another is quite similar to the experiment of mixing two volumes of gas. The latter case can be considered as the free expansions of two volumes of gas into each other (Fig. \ref{freeexp1}, top and bottom).

Consider two initial sub-volumes filled by the same quantity $N/2$ of gas particles, at the same pressure and thermal energy, but which can be of the same species, or of two different species (e.g. two different isotopes). Remove the partition.
Here again, as for the case of free expansion, whether or not an additional work is needed to return the system to the initial state, and so the difference of entropy between the final and initial states, depends on what we know about the initial state. This depends on an additional source of information allowing to distinguish the two gas species and to decide whether or not they are identical.

This experiment of mixing is at the basis of what is known as the Gibbs' paradox of discontinuity of the entropy of mixing: the entropy upon mixing is believed to vary discontinuously from 0 to $N\ln 2$ with the dissemblance whereas other physical quantities vary continuously. 
Gibbs actually considered this as a veridical paradox in the sense that everything is correct (the discontinuity is actual) and there no contradiction\,\cite{Quine1976}. 
He simply noted this discontinuity and immediately gives us the solution: \myquote{It is to states of systems thus incompletely defined that the problems of thermodynamics relate} (J.W. Gibbs \,\cite{Gibbs1874} p.228).


\subsubsection{Incomplete information}

Filled or empty (free expansion), identical or not (mixing), it seems clear-cut (hence the paradox of mixing considered as veridical), but actually it is not the case. The paradox is falsidical and there is no discontinuity.
In the first experiment (free expansion), the actual work needed to restore the initial state depends on how many gas particles between 0 and $N$ must be actually compressed, i.e. what is considered empty for the other compartment. 
In the second experiment (mixing), the actual work depends on which degree of separation of the two species is considered satisfactory.
In both cases, the minimum actual work to return to the initial state varies continuously from 0 to $TN\ln 2$, and thus the entropy difference between the two states from 0 to $N\ln 2$\,\cite{Lairez_2024}.

It could be argued that these are special cases in which the observer does not have full information. Full information does not depend on the observer and could be considered as an intrinsic property of the system that belongs to a Platonic \textit{objective reality}.
But this does not hold logically, because it is impossible, at any moment, to be certain that one has all the information. The total information is and will always remain unknown, and that which we have will always remain incomplete.

\subsubsection{The observer's choice}

The observer suffers the best instrumental resolution and our current common knowledge of the physical world, all things which are not under his control.
But there are other examples where his role is more active and the observer decides for himself which information deserves attention or not.

\begin{figure}[!htbp]
	\begin{center}
		\includegraphics[width=1\linewidth]{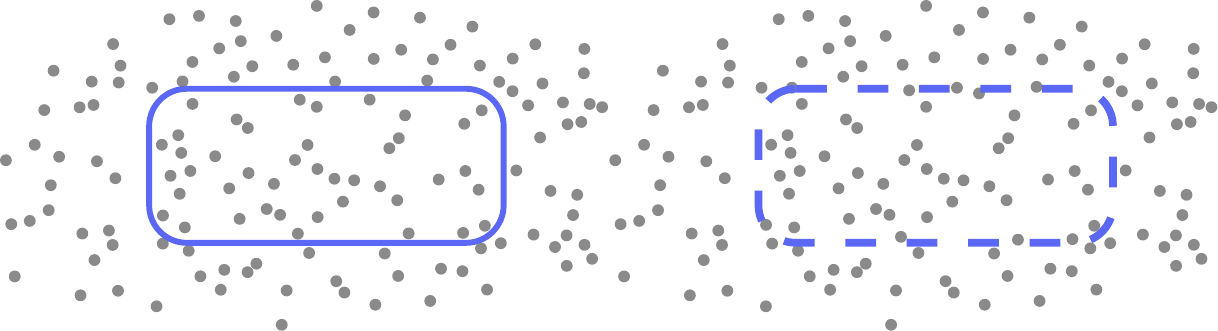}
		\caption{A large reservoir contains a gas. In the left side (red line) a closed sub-volume. In the right side (in dashed red line) an open sub-volume. If the gas particles are distinguishable (traceable) their identity is an information that may contribute to the description of the closed system, but which has no meaning for the open system.}
		\label{ouvert}
	\end{center}
\end{figure}

Consider a sub-volume of gas within a larger reservoir filled with the same gas species made of distinguishable particles (Fig.\,\ref{ouvert}). If the sub-volume is closed, the particles constituting the system are given, so that their identity is information which is not necessarily interesting but which undoubtedly contributes to the total information allowing the system to be described.

On the contrary, if the sub-volume is open, its particles are continually exchanged with the reservoir, so that their identity does not participate at all in the total information about the system, although strictly speaking this information exists. 

With this example, it appears that the observer's definition of the system determines what information is worth considering.
In the thermodynamics of open systems, the identity of the particles is irrelevant, and neither are their permutations. The reason is not that the particles are indistinguishable or untraceable (they are always distinguishable or traceable as soon as they are  large enough). The reason is that we decided that it was not relevant according to the definition we gave of the system.

Let us report a quote from van Kampen: \myquote{The choise between [distinguishable or indistinguishable molecules] depends on whether the experimentalist is able and willing to make the distinction, not on the fundamental properties of the molecules. Thus the [Gibbs] paradox is resolved by replacing the Platonic idea of entropy with an operational definition}\,\cite{van_Kampen_1984}.

\subsection{Information is energy: the proof by demons}

Changes and therefore the energy needed to bring them about are not intrinsic to a process or a system. They also depends on the information we have. Does this mean that information is energy\,? Yes and it is proved by the intervention of demons.

\begin{figure}[!htbp]
	\begin{center}
		\includegraphics[width=1\linewidth]{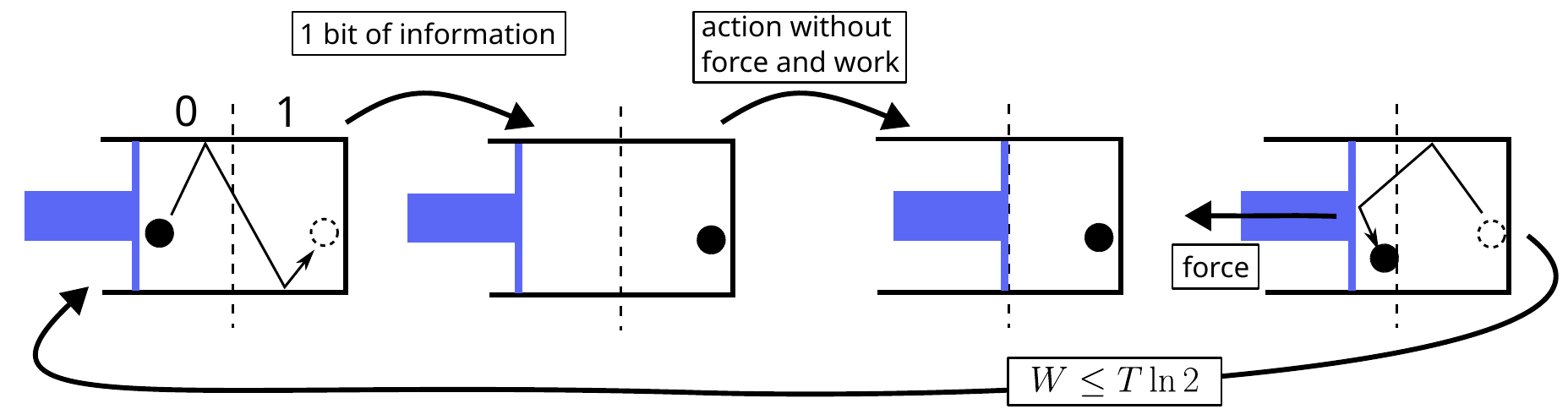}
		\caption{Szilard engine: a demon detects the very moment when the particle is in compartment 1, then he places a piston at no energy cost (other than the information about the position of the particle) allowing to produce a work at best equal to $T\ln2$ from the expansion of the single-particle \textquote{gas}.}
		\label{szilard_demon}
	\end{center}
\end{figure}

From Maxwell\,\cite{Maxwell_1872} to Szilard\,\cite{Szilard_1964} passing by Loschmidt\,\cite{Darrigol_2021} demons observe thermodynamic systems at the scale of particles, acquire information about them, and use this information to produce energy (see Fig.\,\ref{szilard_demon}). Where does this energy come from\,?
It appears to come from nothing but information. So that, according to the conservation principle, information must be added to the open list of known forms of energy. This is how the different forms of energy have always been added to the list\,\cite{Feynmann_Energy}: 
chemical energy, electrical energy, magnetic energy, radiant energy, nuclear energy, potential energy, dark energy... There is absolutely no reason to do otherwise for information. And moreover, doing otherwise would lead to inconsistencies.

Most of the time the action of demons is presented as a very subtle point that would lead us to wrongly believe that the second law is violated (see e.g. \,\cite{Bennett1987, Liboff1997, Leff2002, Kieu2004, Radhakrishnamurty2010, Ciliberto2021, Fontana2022}). Thanks to even more subtle reasoning it is finally not.
But in reality, it is not just the second law that would be challenged by demons. It is the first principle of conservation that would be ruined if information were not considered energy. Fortunately, this is impossible by construction of this principle (see subsection \ref{first_principle}). The principle of conservation necessarily implies a list of different forms of energy that will remain forever open. This solution to the problem posed by demons is certainly less subtle than others, but it is above all more robust.


\section{Statistical mechanics}

Thermodynamics clearly tends towards the Aristotelian conception of \textit{objective reality} and this is precisely where the problem lies for  Platonists. So, since the advent of atoms, attempts have been made to correct this.
This is the reductionist approach of statistical mechanics which aims to derive thermodynamics from atoms, the mechanics of their collisions and from probabilities, in a way more compatible with a Platonic conception.

This presentation may seem biased, it may seem like presumed intent. But it is not and the intent is clearly stated from the beginning by Gibbs himself.
Motivations for reductionism should be the unification with fluid mechanics, or with anything else, but here this is not claimed. Instead, Gibbs who first used the term statistical mechanics in the title of his book\,\cite{Gibbs_1902} subtitles it \textquote{the rational foundation of thermodynamics}. Does this mean that Gibbs considered thermodynamics to be irrational\,? Certainly not. By \textquote{rational} he means \textquote{Platonic}, that is, more objective and more independent of humans than phenomenology. What is more objective and independent of humans than atoms and Newton's laws of mechanics\,? 

But, to realize the program of statistical mechanics, there is a first obstacle to overcome: with which probability distribution should we begin the calculations\,?

\subsection{The pragmatic approach of the fundamental postulate}\label{fundamental_postulate}

With atoms comes thermal agitation. Even at the equilibrium, thermodynamic systems are dynamical: they are continually changing their microscopic configuration (their phase or microstate), which we have absolutely no way of knowing with certainty but only in terms of probabilities. Provided that the distribution is known. This is the first problem to be solved by starting with the simplest case of an isolated system which exchanges neither energy nor matter with the surroundings.

The pragmatic approach to this problem is straightforward:
\myquote{When one does not know anything the answer is simple. One is satisfied with enumerating the possible events and assigning equal probabilities to them.} (R. Balian\,\cite{Balian_1991} p.143).
This is the fundamental postulate of statistical mechanics, that is a variation of the Laplace's \textquote{principle of insufficient reason}\,\cite{Dubs_1942} or \textquote{principle of indifference}\,\cite{Keynes_1921}: basically, all microscopic configurations have same probability because we have no reason (no information) to think otherwise. Probabilities are viewed as \textquote{expectations} that depend on the information we have.

However, this solution is felt to be intellectually unsatisfying. Probably the strongest argument against this approach is this:
\myquote{It cannot be that because we are ignorant of the matter we know something about it.} (R.L. Ellis\,\cite{Ellis_1850}), which is very difficult to contest (except in the framework of information theory that will be the subject of the next section).
Thus, the alternative solution retained by Platonists is that of the ergodic hypothesis: probabilities are viewed as frequencies of occurrence of configurations and are intrinsic to the system.

\subsection{The ergodic hypothesis}\label{ergodic_hypothesis}

Consider like Gibbs (\cite{Gibbs_1902} p.1) a Hamiltonian mechanical system made of $N=n/3$ colliding particles (so that $n$ is the degree of freedom), each requiring to be fully described only three coordinates for its position $r$ and three others for its momentum $p$. A microscopic configuration, also called a phase or microstate, is thus a point of coordinates $\{r_1, ,\dots r_{n}, p_1,...p_n\}$, abbreviated $\{\textbf r, \textbf p\}$, in a $6N$-dimensional space, namely the phase space.
Even in the macroscopic stationary state of equilibrium,
the thermal agitation causes the system to continually change phase, that is, to move along a continuous trajectory in phase-space. Let us look at some properties of this trajectory.

In a continuum, the probability of a point $\{\textbf r, \textbf p\}$ has no meaning but must rather concern a given volume around that point $\diff^{3N}\textbf r \diff^{3N}\textbf p$ and involve a local volume-density of points $\rho(\{\textbf r, \textbf p\}, t)$. This density is up to a normalization factor a density of probability and once multiplied by the volume gives the number of phases within this volume.
If the system is isolated, its total energy (its Hamiltonian) is strictly constant and each phase-point of the elementary volume in question must strictly fulfill this constraint. So that, the considered elementary-volume belongs actually to the subset $\Gamma$ of the phase space that includes only all phases that are compatible with the constraint of a given constant energy, that is to say the space of all compatible phases (a ($6N-1$)-dimensional hyper-surface of the phase space). $\Gamma$ is the space of possible phases.

Due to dynamics, according to the different trajectories of each phase-point, any elementary volume will move and evolve within $\Gamma$, just like an elementary volume of a fluid under a flow.
A strictly constant total energy means no fluctuation, so that the flowing fluid of phase-points is actually incompressible: its density is strictly independent of time and position on the trajectory.
\begin{equation}\label{volume_preserving}
	\rho(\{\textbf r, \textbf p\}, t) = \text{cst}
\end{equation}
This is the Liouville's theorem. The density of probabilities of phases throughout a trajectory in $\Gamma$ is uniform. The dynamics is said to be volume-preserving.

Consider two trajectories, since the system is deterministic, they cannot intersect, otherwise at the intersection the system evolution would not be determined. Two trajectories are either disjoint or identical. The ergodic hypothesis is that there are identical. Thus, if we prepare two samples with exactly the same macroscopic constraints (therefore two macroscopically identical samples) but at time $t_0$ two supposedly different phases, the trajectories of the latter will be the same, except that they will have started at a different phase-point. Therefore, this only trajectory encompasses all possible phases for an isolated system.
In other words, the set of phase-points of this trajectory and the ensemble $\Gamma$ of possibilities, namely the microcanonical ensemble, are only one. The density of probabilities of phases in the microcanonical ensemble $\Gamma$ is thus uniform.

\subsection{The link with thermodynamics}

Once the probability distribution for an isolated system is established, thanks to the ergodic hypothesis, other cases can be derived with much less strong hypotheses, and in particular the canonical case of a closed system in contact with a temperature reservoir\,\cite{Lairez_2022a}. With these distributions, the next step of statistical mechanics would be to calculate and express other quantities, particularly that related to the evolution, or absence of evolution, of a system, namely its entropy. 
In other words, starting only from the atomic scale and probabilities, the questions are: 
\begin{enumerate}
	\item Is there a way, free from any \textit{a priori} connection to thermodynamics, to derive a formula for entropy that allows us to define it\,? 
	\item Is there a way, free from any \textit{a priori} connection to thermodynamics, to predict the time evolution of a dynamical system of particles\,? 
	\item Is there a way to define equilibrium not only as stationary (as in thermodynamics) but also as the stable state\,?
\end{enumerate}
These three questions are closely related and it is important to realize that they are absent from phenomenological thermodynamics, but only introduced by statistical mechanics.
1)~In phenomenological thermodynamics, there is no need of a formula to calculate entropy, because it is a measurable quantity (it is the heat exchanged for a reversible change). Measurement makes definition. 2)~Phenomenological thermodynamics deals with states, which by definition are static. It deals with changes of state, it does not deal with what happens during the change but only with what the state was before and what it is after. Dynamics and time variable were introduced by statistical mechanics.
3)~In phenomenological thermodynamics, the equilibrium is stationary. Once at equilibrium, an isolated system remains there because nothing is supposed to move it away from it. Statistical mechanics introduces probabilities and therefore fluctuations into the theory. The system may spontaneously deviate from equilibrium. So that, the stationary state must be stabilized by something that must also be introduced into the theory.

In this section, these three questions are detailed with particular attention to the consistency of the answers they receive and to the reality of the emancipation from thermodynamics that they promise.

\subsubsection{Gibbs and Boltzmann Entropies}

\begin{figure}[!htbp]
	\begin{center}
		\includegraphics[width=0.8\linewidth]{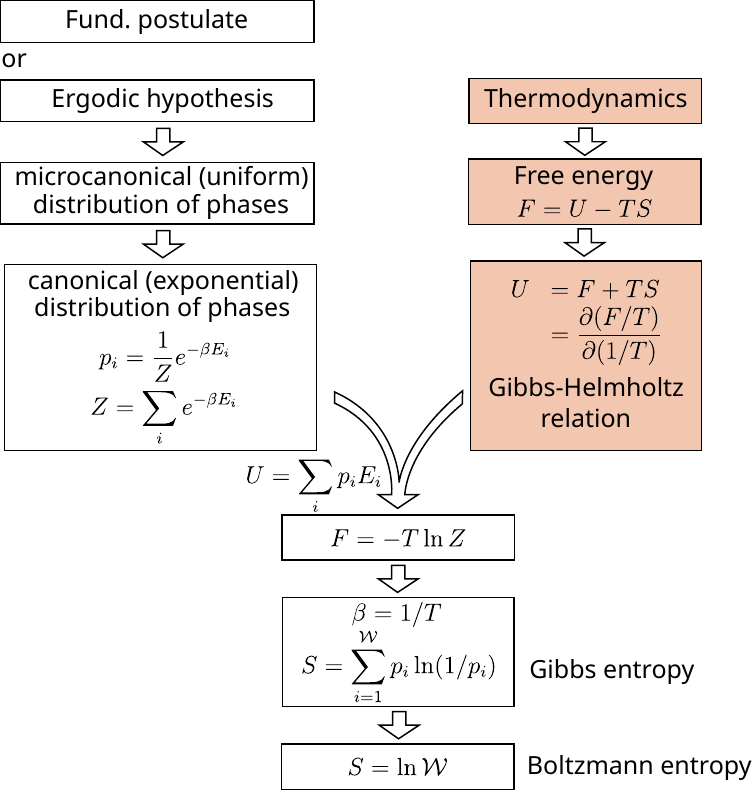}
		\caption{An example of derivation of Gibbs and Boltzmann entropy formulas in statistical mechanics. The derivation is never totally free of thermodynamics results and formulas. Details of the derivation can be found in ref.\,\cite{Lairez_2022a}}
		\label{link_thermo}
	\end{center}
\end{figure}

Is there a way, free from any \textit{a priori} connection to thermodynamics, to derive a formula for entropy that allows us to define it\,? 
The answer is no. At some point or another, thermodynamics will come into play in the derivation. Figure \ref{link_thermo} gives the outline of an example of derivation\,\cite{Lairez_2022a}, other examples of derivation can be found in ref.\,\cite{Tien_1979,Sekerka_2015}.
All methods amount to calculating from the distribution of phases certain statistical quantities, which by identification with classical equations of thermodynamics, lead to the Gibbs' statistical entropy formula for closed systems:
\begin{equation}\label{GS}
	S_\text{G}=\sum\limits_{i}^{\mathcal W} y_i\ln 1/y_i,
\end{equation} 
or to that of Boltzmann for isolated systems:
\begin{equation}\label{BS}
	S_\text{B}=\ln{\mathcal W}
\end{equation} 
where $\mathcal W$ is the cardinality of the discretized space $\Gamma$ of possible phases and $y_i$ the probability of phase $i$.

Compared to the initial ambition of statistical mechanics, the result is a little disappointing.

\subsubsection{Towards equilibrium and H-theorem}\label{H_theorem}

Is there a way, free from any \textit{a priori} connection to thermodynamics, to predict the time evolution of a dynamical system of particles\,? 
This question is implicitly related to entropy and to the Clausius inequality (Eq.\,\ref{Clausius_ineq1}). 
If a spontaneous change of an isolated system towards an equilibrium state corresponds in the end to a positive change in entropy, it seems natural that for a dynamical system of particles there must exist a time-dependent statistical quantity that increases with time during the change.
Boltzmann addressed this question of evolution towards equilibrium and looked for such a probability rule that would substitute to the second law of thermodynamics\,\cite{Boltzmann_Lectures}.

Starting from the kinetics theory of gases of Maxwell\,\cite{Maxwell_1867} and in particular from the distribution of velocities he obtained, Boltzmann approached the problem of collision between two particles also in terms of probabilities. Because, even if the result of a given collision is purely deterministic, the exact conditions in which this collision occurs is unpredictable and random. This is the hypothesis of \textquote{molecular chaos}.
For his calculation\,\cite{Boltzmann_Lectures, Cercignani_1988}, Boltzmann considers the 6-dimensional phase-space $\Gamma_1$ of one single particle and denotes $f(t)$ the corresponding probability density at time $t$. From the hypothesis of molecular chaos, he obtains a transport equation that allows him to argue that a certain statistical quantity $H(t)$ always increases with time until it becomes constant at the equilibrium\,:
\begin{equation}\label{H2}
	\frac{\diff H}{\diff t}\ge 0
\end{equation}
with
\begin{equation}\label{defH}
	H(t) 
	= \int_{\Gamma_1} \diff^3\textbf{r}\diff^3\textbf{p} \quad f(t)\ln (1/f(t) )
\end{equation}
These equations forms the H-theorem.
It was the first time that anything resembling statistical entropy had been written. The H-theorem seems natural and is frequently viewed\,\cite{Villani_2008, Weaver_2021, Weaver_2022} as an alternative statistical expression of the 2nd law of thermodynamics plus a definition of equilibrium (equality sign in Eq.\,\ref{H2}) that ensures its stability.

Except that the mathematical proof of this theorem is still in progress\,\cite{Cercignani_1988, Villani_2002}. But moreover, except that $H$ is not yet entropy. Because here, $f$ is a density of particle that differs from a density of probability of phase, so that it is not yet related to thermodynamic measurable quantities. 
Thus, even if it may be interesting to avoid any \textit{a priori} link with thermodynamics in the derivation of the 2nd law (and in this sense the H-theorem is a great success), the absence of an \textit{a posteriori} link is problematic.

But let us leave these points aside and focus on another problem that was raised by Loschmidt\,\cite{Uffink_2006,Darrigol_2021}. Newton's equations of classical mechanics do not introduce any time irreversibility. If a motion is possible in one direction, it is also possible in the opposite direction. 
Loschmidt imagines a dynamical system made of colliding particles that tends towards equilibrium. At a given moment, a demons is capable of reversing the velocities of all the particles. 
This new microscopic state is just as compatible as the previous one with the macroscopic properties of the system, but the latter will now evolve backward and deviate from equilibrium. This is incompatible with the H-theorem. So, we wonder by what sleight of hand does Boltzmann arrive at his result\,? So, we wonder how irreversibility suddenly arises\,?
It is actually put from the beginning of the reasoning under the form of the hypothesis of \textquote{molecular chaos}. A collision is conceived as fundamentally asymmetric in time: random before, deterministic after.
This is acknowledged by Boltzmann himself: \myquote{Any attempt to deduce [the Clausius inequality] from the nature of the bodies and from the law for the forces acting among them, without invoking the initial conditions, must therefore be in vain}\,\cite{Darrigol_2021}. 

The H-theorem is obtained because the 2nd law has been moved to the root of proof. Therefore, considering the H-theorem as a proof of the 2nd law would clearly be circular reasoning.

\subsubsection{The paradox of the stability of equilibrium}

Probabilities introduce fluctuations. A system at equilibrium is stationary on average, but may transiently deviate from equilibrium. It is therefore imperative to introduce something into the theory, an additional postulate or principle, that would play the role of an internal restoring force. Otherwise, the equilibrium would theoretically not be stable.
In mechanics, a body will spontaneously follow a descent path of potential energy until it reaches the minimum.
Similarly, in thermodynamics, we already have the 2nd law, according to which the entropy of an isolated system can only increase spontaneously. The postulate that entropy is maximal at equilibrium would ensure stability\,\cite{Callen_1985}.
This postulate would make the equilibrium an attractor.

Here is the main problem. An attractor is a region of the phase space where many trajectories will converge, so that the local phase-density will increase. And this is incompatible with a volume-preserving dynamics and a constant probability density throughout the trajectory (Eq.\,\ref{volume_preserving}).


This contradiction is particularly apparent in light of Poincaré's recurrence theorem, which states that any volume-preserving dynamical system, whose trajectory is restricted to a bounded region of phase space (for instance with a finite volume and given total energy), behaves recurrently with arbitrary precision.
In short, the system always returns at some time or another as close as desired to any phase already visited.
This recurrence actually justifies considering probabilities as frequencies, but it leads to expected behaviors of the system radically opposed to observations and radically opposed to the conception of equilibrium as being stable.
This is the paradox mentioned by Zermelo: it cannot be excluded that all the gas particles from an open bottle that have escaped into the room will concentrate themselves back into the bottle. 

The argument for resolving this paradox\,\cite{Boltzmann_Lectures, Villani_2012} is based on different time scales: that of observations which is very short compared to that of recurrence which is geological (so that its probability tends to zero).
The latter argument resolves the apparent contradiction that can be seen in the fact that something that is possible in principle is never observed in practice.
But this does not resolve the real contradiction of considering a possibility that never occurs as an intrinsic property of a system.
This does not resolve the real contradiction of considering probabilities of events as frequencies while denying their recurrence.
This does not resolve the real contradiction of considering equilibrium as stable while not being an attractor.

\subsection{Gibbs paradox of joining two volumes of gas}\label{Gibbs_paradox2}

A second paradox is introduced by statistical mechanics. It is related to the Gibbs paradox of mixing evoked in \S\ref{Gibbs_paradox1}, so that it is often called the second Gibbs paradox, even though Gibbs never mentioned it. 

The first Gibbs paradox concerns the joining of two volumes of gas consisting of different or identical species. The second paradox concerns only the latter case, but considered from the point of view of thermodynamics or that of statistical mechanics. The removal of the partition between two identical volumes $V$, filled with the same gas at the same temperature and pressure, is not accompanied by any measurable exchange of heat or work with the surroundings. When the partition is put back in position, the system is restored to its original state.
The difference of Clausius entropy between the joined and disjoined states is thus zero:
\begin{equation}\label{Delta_Clausius}
	\Delta S=0
\end{equation}
From the point of view of phases probabilities, it is not so simple. In equation \ref{BS}, let us write $\mathcal W$ as:
\begin{equation}
	\mathcal W = \mathcal R \times \mathcal P
\end{equation} 
where $\mathcal R$ is the number of possibilities in position for $N$ particles and $\mathcal P$ the number of possibilities in momentum. 
For a volume $V$ with $N/2$ particles,
the number of possibilities in position is $V^{N/2}$. Thus for two equal disjoined volumes $V$ with $N/2$ particles each, one has:
\begin{equation}\label{R_disjoin}
	\mathcal R_\text{disjoined} =  V^{N/2} \times V^{N/2} = V^N
\end{equation}
Once the partition is removed, the position of each particle can now be anywhere in $2V$ (instead of anywhere in~$V$). So that $\mathcal R$ becomes $(2V)^{N/2} \times (2V)^{N/2}$, or:
\begin{equation}\label{R_join}
	\mathcal R_\text{joined} =  (2V)^{N} = 2^N \times V^N
\end{equation}
In both cases, disjoined and joined, the distribution of particles momentum is the same (the total energy is unchanged).
Thus, the difference in Boltzmann entropy between the two states is $\Delta S_\text{B}=\ln \mathcal R_\text{joined}/\mathcal R_\text{disjoined}$, leading to:
\begin{equation}\label{Delta_Bolzmann0}
	\Delta S_\text{B}= N\ln 2
\end{equation}
To be compared to Eq.\,\ref{Delta_Clausius}. The puzzle is that both entropy variations should be the same since Boltzmann's entropy is derived from Clausius's (Fig.\,\ref{link_thermo}). 

To this paradox, usually two solutions are proposed\,\cite{Ray_1984,Huang_1987, Nagle_2004, Cheng_2009, Versteegh2011, Frenkel_2014, Dieks_2018, Saunders_2018, van_Lith_2018}) (except the one involving the notion of information presented in the next section)\,:

\begin{itemize}
\item \textbf{Quantum mechanics solution}: according to this point of view it is not possible to understand classically the difference between Eq.\,\ref{Delta_Clausius} and \ref{Delta_Bolzmann0}, the solution is quantum (\cite{Huang_1987} p.141).
The number of possibilities for the joined and disjoined states are both overestimated because identical particles cannot be distinguished. In the disjoined case, the $(N/2)!$ permutations of a combination of $N/2$ particle-positions must be counted as 1. Eq.\,\ref{R_disjoin} must be replaced by:
\begin{equation}\label{R_disjoin_correct}
	\begin{array}{rl}
	\mathcal R_\text{disjoined} &\displaystyle = \frac{ V^{N/2}}{(N/2)!} \times \frac{ V^{N/2}}{(N/2)!}\\ \\
	&\displaystyle= V^N\times \frac{1}{(N/2)!\times (N/2)!}	
	\end{array}
\end{equation}
and similarly, Eq.\,\ref{R_join} must be replaced by:
\begin{equation}
	\mathcal R_\text{joined} =   {2^N}  V^N\times \frac{1}{N!}
\end{equation}
So that these corrections give:
\begin{equation}\label{correct_quantum}
	\Delta S_\text{B}= N\ln 2 - \ln \frac{N!}{(N/2)!(N/2)!}
\end{equation}
instead of equation \ref{Delta_Bolzmann0}. 
This argument is known as \textquote{correct} Boltzmann counting, even though it is no more correct than the following one.
		
\item \textbf{Classical mechanics solution}: The number of possibilities in the disjoined state is actually {underestimated} in Eq.\,\ref{R_disjoin}\,\cite{Ray_1984, Cheng_2009, Versteegh2011}. Because when partitioning, a particle can end up in either of the two compartments leading to as much possibilities which are not counted in Eq.\,\ref{R_disjoin}. The correct calculation should be increased by a factor accounting for the different possible combinations of partitioning $N$ particles in two sets of $N/2$ each:
\begin{equation}\label{R_disjoin2}
	\mathcal R_\text{disjoined} =   V^N \frac{N!}{(N/2)!\times (N/2)!}	
\end{equation}
So that, using Eq.\,\ref{R_join} and \ref{R_disjoin2} one obtains for the  difference of Boltzmann entropy:
\begin{equation}\label{correct_classic}
	\Delta S_\text{B}= N\ln 2 - \ln \frac{N!}{(N/2)!(N/2)!}
\end{equation}
that is exactly the same as Eq.\,\ref{correct_quantum} but with a completely different argument.
\end{itemize}	

These two solutions in combination with an approximation of the Stirling formula in  Eq.\,\ref{correct_quantum} or \ref{correct_classic} for the calculation of the second term give $\Delta S_\text{B}=0$.
The paradox is supposed to be resolved. In reality, as it stands, it is not, and it can be shown that these solutions benefit from the cancellation of two approximations\,\cite{Lairez_Stirling}. The solution must be amended by considering fluctuations and using of the exact Stirling formula (rather than an approximation of it).
But the purpose of this article is rather to highlight what exactly these solutions entail regarding the role of the observer.

The first point concerns the \textquote{undistinguishability} of particles. Actually, as soon as a particle is large enough to be traceable by the mean, for instance, of a microscope and a camera, it becomes distinguishable from the others. Single-particle tracking is now feasible for particles above 20\,nm (see for a recent review \cite{Kobayashi2025}). The implementation to many particles is a question of computing power and there is no conceptual impediment to this (for recent papers on these technics see e.g. \cite{Crocker2007, Pinto2021, Xu2024}). The question is therefore not whether the particles are distinguishable or not, but whether they are distinguished or not.
The difference is very important, because in the second case it depends on the observer. It depends on the means and information available to the observer. And it depends on what the observer considers to be relevant. Of course, before the age of atoms, thermodynamics did not care about tracking and distinguishing particles, because it simply did not care about particles. Despite this, the theory was and still is correct.


The second point concerns the solution proposed in the framework of classical mechanics.
This time, particles can be distinguishable and distinguished, but in counting the number of possibilities offered to the system, we add the different possible combinations for the distribution of the particles between the two compartments.
There is no problem with this, except that it must be understood that once the partition is established in the disjoined state, there is no possibility for the system to modify it. So these possibilities are not possibilities accessible to the partitioned system at equilibrium, they are possibilities accessible just before the partition is carried out. On this respect, the partitioned system at equilibrium (the disjoined system) is no longer ergodic\,\cite{Peters_2013}. This solution is actually incompatible with the ergodic hypothesis. The different combinations are not possibilities offered to the system, they simply contribute to the uncertainty that the observer has about it.
	
\section{Shannon's information theory}

Shannon's information theory\,\cite{Shannon_1948} initially aimed to optimize physical media for the storage (optimization in terms of size) and transmission (optimization in terms of bandwidth) of information conceived as the random outcomes of a source. 
This concrete physical problem seemed far removed from those of thermodynamics. However, it turned out that this was not the case.

\subsection{Quantity of information}\label{Quantity_info}

Consider a source that sequentially emits a message whose letters from the receiver's point of view are as many random outcomes $\vx$, which have to be encoded and stored.
For the encoding, the first thing to do is a mapping $\mathcal m$ between the finite space $\Gamma$ of  possibilities for $\vx$ and a subset of natural numbers:
$$
	\begin{array}{llll}
	\mathcal m: & \vx\in\Gamma & \longmapsto & n\in\{1,..\W\} \subset	\mathbb{N}
\end{array}
$$
With this mapping, since $m(\vx)=n=2^{\log_2 n}$, $\vx$ requires $\log_2(n)$ bits to be recorded. Thus, a first approach to the problem is to plan a fixed length equal to $\log_2(\mathcal W)$ for each outcome to be sure not to lose information\,\cite{Hartley_1928}. But this is clearly not always optimal (see Table \ref{table_code}) as it introduces for small numbers unnecessary 0 at the beginning of their binary code. An encoding procedure that uses just the necessary number of bits $\log_2(\mathcal n)$ for each outcome $n$ is much more optimal.

\begin{table}[h]\label{table_code}
	\begin{center}
	\begin{tabular}{c|c}
		Fixed length encoding & Variable length encoding\\
		\hline
		$\begin{array}{r}
			0001 0001\\
			0000 0001\\
			0000 0101\\
		\end{array}$ &
		$\begin{array}{r}
			1 0001\\
			1\\
			101\\
		\end{array}$\\
	\end{tabular}
	\caption{Binary encoding of random outcomes $n=\mathcal m(\vx)$. Left: fixed length encoding (here 8 bits per outcome) introduces unnecessary 0 for small $n$. Right: variable length encoding saves storage space.}
\end{center}
\end{table}	

However, the storage space saved with a variable-length encoding rule depends largely on the mapping chosen. To optimize storage space, rare outcomes should be mapped to large numbers while frequent outcomes should be mapped to smaller numbers.
Shannon\,\cite{Shannon_1948} showed that in no case, however, can the minimum average number $H$ of bits per outcome be less than
\begin{equation}\label{Q_information}
	H(p) = \sum\limits_{\vx\in\Gamma}  p(\vx) \log_2(1/p(\vx))
\end{equation}
By writing
\begin{equation}\label{Shannon_entropy}
S_\text{S}(p)=H(p)\ln 2	
\end{equation}
we obtain the Shannon entropy of the dynamical source. $H$ is the minimum average number of bits per outcome needed to encode the random outcomes $\vx$ of the source.
$H$ is called the quantity of information emitted by the source. It is the information we would need to describe the source, understand an information we do not necessarily have.

Here, let us focus on the meaning of the word \textquote{information} in Shannon's theory. 
Imagine that $\vx$ is a random natural number lying in a given interval. Denote $H_0$ the quantity of information of the source given by Eq.\,\ref{Q_information}.
Imagine you learn, by one way or another, that the outcomes are all even. This means that you know in advance that the first bit of all binary encoded outcomes is~0. So you do not need to store it every time. You have saved one bit of information:  $H=H_0-1$.
An important point is that the latter information we got about the outcomes distribution remains valid until the parity changes. It is not needed to give us this information twice. In fact, the same information cannot be given twice. The second time, it is no longer information.
When we get information from reading a newspaper, the amount of information we have learned does not increase if we read the same newspaper a second time.
The meaning of the word information in Shannon's theory is in fact the usual meaning of the word. Shannon only found a way to quantify the information.
This will be a key point in what follows.

Shannon entropy (Eq.\,\ref{Shannon_entropy}) is nothing other than the Gibbs formula (Eq.\,\ref{GS}) in the case where the outcomes considered are the phases of a dynamical system. 
With this result, the Gibbs formula is obtained without any connection to thermodynamics. It is interesting, but for now it only seems like an analogy. 
Just as with the H-theorem (see \S\ref{H_theorem}), a connection must be established for this analogy to be of any use in statistical mechanics. But the difference is that here, it can be established without inconsistency. 
The connection begins with the maximum-entropy theorem and continues with its application by Jaynes\,\cite{Jaynes_1957} to the problem of statistical mechanics that led to the maximum entropy principle.

\subsection{Maximum Shannon entropy theorem (MaxEnt)}

Statistical mechanics faces the problem of determining from which distribution of phases to start calculations. Hence the fundamental postulate (section \ref{fundamental_postulate}) or the ergodic hypothesis (section \ref{ergodic_hypothesis}) options.
The approach of information theory to solving this problem is the one commonly used in modeling any set of experimental measurements. We must take into account all available information, but no more. This is typically achieved by maximizing uncertainty within the constraints of satisfying our knowledge.

In his paper\,\cite{Shannon_1948}, Shannon showed that the quantity of information $H(p)$ (Eq.\,\ref{Q_information}) is actually the only measure of uncertainty on $\vx$ that combines three interesting qualities: 1)~continuous in $p$; 2)~increasing in $\mathcal W=1/p$ for a uniform distribution; 3)~additive over different independent sources of uncertainty.
In addition, for the problem of seeking for the probability distribution of a random variable, Shore and Johnson\,\cite{Shore_1980} showed that maximizing $H(p)$ is the only procedure allowing to obtain a single solution for the probability distribution. Hence the theorem:

\law{Maximum Shannon entropy theorem (MaxEnt)}{\\The best distribution $p(\vx)$ that maximizes the uncertainty on $\vx$, while accounting for our knowledge, is the one that maximizes Shannon entropy.}	

The \textit{a priori} probability distribution of anything can thus be determined with a rational criterion. \textit{A fortiori} the distribution of phases of an isolated system. The distribution with which to start the calculations of statistical mechanics can thus be determined rigorously with the method of Lagrange multipliers where the imposed constraints are the knowledge we have about the system. The result is exactly the same as that of the fundamental postulate or that of the ergodic hypothesis: a uniform probability distribution of phases for an isolated system.

\subsection{Maximum entropy principle}

The maximum Shannon entropy theorem provides a rational criterion to determine the prior probability distribution of a random variable. But this does not provide a criterion that could serve as a definition of equilibrium. This does not tell us which random variable would have its distribution maximizing Shannon entropy at equilibrium. This does not legitimate the use of phase distribution. Jaynes\,\cite{Jaynes_1973} noted that, based on the basic postulate that equilibrium is unique, the variables to be considered must have a distribution that would be invariant in form under similarities (rotation, translation and scaling), like the distribution of phases, but also like the number density of particles.
With this postulate, plus the MaxEnt theorem, we can write a principle on which to base statistical mechanics:

\law{Maximum Shannon entropy principle}{\\The equilibrium of a system is the only state that maximizes Shannon entropy of variables whose distributions are similarity-invariant in form.}

\noindent This principle, together with the 2nd law, makes the equilibrium an attractor.

\subsection{Representationalism}

Shannon entropy is a measure of uncertainty. So, there is one question in relation with the previous definition of equilibrium that may bother some. Why would a system at equilibrium (an object) maximize the uncertainty we (subjects) have about it? This would be like a role reversal.

Actually, there is no role reversal because with information theory applied to statistical mechanics, as in thermodynamics, there is no dualism. The observer is present from the outset in the problem to be solved. The maximum Shannon entropy principle can be paraphrased as follows: 1)~All knowledge of equilibrium originates
from observations; 2)~This knowledge allows us to make a mental representation of the equilibrium; 3)~This mental representation, to be rational, must maximize the uncertainty, so as not to invent knowledge that we do not have. In the statement of the above maximum entropy principle, \textquote{equilibrium} has to be thought as \textquote{the mental representation of the equilibrium}.

Let us take an example. Imagine we are rolling a die for a public game and are tasked with deciding the fair probabilities (the odds) of each outcome. In a first case, someone tells us that the die is normal, so that the probability for each outcome is 1/6 (uniform distribution). In a second case, someone tells us that the die is loaded but does not tell us how. So, we have absolutely no other rational choice but to leave the previous odds for betting unchanged (even though you know it is wrong), until the actual probability distribution has been measured. But this measurement is not possible for phases.

The probability distribution of anything, which is to be rationally used as a basis for further calculations, is not an intrinsic property of the system. It is a property of the mental representation we have of it. 
So that, the object of the calculations in question is this mental representation. Subject and object merge.
	
\subsection{Statistical mechanics becomes\\ probabilistic mechanics}
	
The maximum Shannon entropy principle makes the equilibrium stable. 
All right, but what is the difference with establishing directly this principle without information theory\,? 

Without information theory, we would actually need two postulates: 1)~the fundamental postulate (\ref{fundamental_postulate}) or the ergodic hypothesis (\ref{ergodic_hypothesis}); and 2)~the maximum entropy principle. The first to determine the distribution of phases at equilibrium, the second to stabilize equilibrium.
Beyond the economy provided by information theory (one principle instead of two), the latter brings the consistency that was absent before.
The MaxEnt theorem does not suffer from the criticisms that can be made against the principle of insufficient reason (fundamental postulate).
Also, the maximum Shannon entropy principle does not suffer from the inconsistency of the ergodic hypothesis. Because with information theory, probabilities are expectations that depend on our knowledge (\textit{a priori} probabilities), they are not frequencies (\textit{a posteriori} probabilities which are supposed to be measurable but cannot be measured, probabilities which are supposed to be frequencies but concern outcomes that are not recurrent in practice).

For the rest, by starting with the maximum Shannon entropy principle, everything in statistical mechanics is the same, but without dubious starting assumption and
with the added bonus of solving some puzzling questions: the 2nd Gibbs paradox and the puzzles posed by demons.

\subsection{Gibbs paradoxes}

The first Gibbs paradox (section \ref{Gibbs_paradox1}) concerning the mixing of identical or different gas species
was actually already solved by Gibbs himself: \myquote{It is to states of systems thus incompletely defined that the problems of thermodynamics relate} (J.W. Gibbs \,\cite{Gibbs1874} p.228). The second Gibbs paradox (section \ref{Gibbs_paradox2}) concerning the joining of two volumes of the same gas species, was introduced by the Boltzmann formula for entropy. With information theory, we find a solution for the second paradox that is in line with that of Gibbs for the first\,:
\begin{itemize}
	\item Clausius entropy (Eq.\,\ref{Clausius_definition}) concerns a system about which we have incomplete information.
	\item Boltzmann (or Gibbs) statistical entropy (Eq.\,\ref{GS} and \ref{BS}) concerns a system about which we are supposed to have all the information.
	\item Shannon statistical entropy (Eq.\,\ref{Q_information} and \ref{Shannon_entropy}) gives the same formula than Boltzmann or Gibbs, but probabilities have not the same meaning. They allows us to account for the actual information we have, or the information we consider relevant \,\cite{Lairez_2024}.
\end{itemize}

With information theory, we can consider fluctuations or not, isotope composition or not, impurities or not etc. and we can obtain all intermediate cases between that of thermodynamics and that of a fully known system on which we have all the information.
But how to be sure to have all the information without closing the door to further progress in knowledge\,? Information theory avoids this paradox.

\subsection{Brillouin negentropie principle of information and demons}
\label{reformulation_Clausius}

The general formula for Shannon entropy makes the Gibbs formula (Eq.\,\ref{GS}) a special case.
But Shannon's result does not invalidate the connection between Gibbs and Clausius entropy (Fig.\,\ref{link_thermo}) that has been previously established.
Simply, this connection is made \textit{a posteriori} since it was not used for the derivation of the formula. Anyway, Clausius entropy $S$ and Shannon quantity of information are linked, via Gibbs entropy and this allows us to write:
\begin{equation}
	 S=H_{\text{phases}}\ln 2
\end{equation}
with $H$ given by Eq.\,\ref{Q_information}. It follows that the Clausius inequality (Eq.\,\ref{Clausius_ineq2}) 
can be rewritten as:
\begin{equation}\label{Clausius_Shannon}
W\ge - T\Delta H_{\text{phases}} \ln 2	
\end{equation}
where $H_{\text{phases}}$ is the uncertainty on the actual phases of the system.

Acquiring any information about the current phase of a dynamical system reduces the uncertainty we have about it.
Or in other words, the information we get about a system has a negative contribution to its entropy\,\cite{Brillouin1953, Brillouin_1956_book}.
For one bit of acquired information, we have $\Delta H=-1$. By using Eq.\,\ref{Clausius_Shannon}, one gets:
\begin{equation}\label{wacq}
	W_\text{acq/bit} \ge T \ln 2
\end{equation}
where $W_\text{acq/bit}$ is the work that we have to provide (and that will be dissipated as heat) per bit of acquired information. Brillouin called this the \myquote{Negentropie principle of information} \cite{Brillouin1953, Brillouin_1956_book}.
This is in fact nothing more than a reformulation of the Clausius inequality that takes into account Gibbs' and Shannon's mathematical results connecting Clausius entropy to statistical entropy and finally to the quantity of information.

With equation \ref{wacq}, the acquisition of information has a cost, so that demons are no longer demoniac.

Acquiring information about a system is a complex process consisting in many elementary operations which have to be cyclically performed if the system is dynamical. These operations include measuring, writing the measurement result, but also copying, overwriting, reading, erasing memory etc. 
Equation \ref{wacq} tells us about the minimum work to be provided, or the minimal heat dissipated, at the end of one cycle for one bit of information. This equation concerns the net energy balance of the global cycle.
It is general without exception, as is the 2nd law.
But Eq.\,\ref{wacq} does not tell us precisely where, in the cycle of elementary operations, the inevitable dissipation occurs. It does not tell us which step causes this dissipation. Moreover, this equation does not tell us that such a particular dissipative step exists. 
There is no obligation for this, it is the entire cycle that dissipates a minimum of heat, not any particular step.
The \textquote{location} of dissipation (if it is localized) will depend on each specific case: which system is considered, which information, which treatment exactly, etc.
It is often mentioned (e.g. \cite{Ciliberto_2018a}) that Brillouin locates dissipation in the measurement stage. This is true. But this only occurs after the statement of the principle of negentropy\,\cite{Brillouin_1956_book}. This principle remains free of any constraints regarding the location of the dissipation.

\section{Landauer erasure principle}

It is not possible to deny the advantages of introducing the notion of information in statistical mechanics. And in reality no ones does. The criticism is situated at an epistemological level:
\myquote{[This approach] is associated with a philosophical position in which statistical mechanics is regarded as a form of statistical inference rather than as a description of objective physical reality} (Penrose\,\cite{Penrose_1979}).
\myquote{Such a view, [...] would create some profound philosophical problems and would tend to undermine the objectivity of the scientific enterprise} (Denbigh \& Denbigh\,\cite{Denbigh_1985}).
\myquote{It is an approach which is mathematically faultless, however, you must be prepared to accept the anthropomorphic nature of entropy} (Lavis\,\cite{Lavis_2015}).
It could not be clearer that the problem lies in the conception of \textit{objective reality} which is not Platonic enough.

For Platonists, the relationship between energy and information would be much more acceptable if we had something that would allow us to affirm that \myquote{information is physical}\,\cite{Landauer_1991}. \textquote{Physical}, translate \textquote{independent of us}, \textquote{objectively materialized by something}.
This is the program of Landauer\,\cite{Landauer_1961, Landauer1996}, followed by Bennett\,\cite{Bennett_1982, Bennett_2003}, who tackled the problem via that of demons.

Acquired information must be stored, at least temporarily (otherwise it is not acquired), under the form of data-bit in memory. Thus, the finite size of the memory of any concrete implementation of a demoniac machine necessarily implies a cyclic erasure of data-bit.
The Landauer's claim is that the unavoidable heat dissipation for one bit of information (Eq.\,\ref{wacq}) discussed in \S\ref{reformulation_Clausius}, is precisely due to this erasure.
\myquote{Landauer’s principle [...] makes it clear that information processing and acquisition have no intrinsic, irreducible thermodynamic cost whereas the seemingly humble act of information destruction does have a cost} (Bennett\,\cite{Bennett_2003}).

In other words, Landauer erasure principle, by precisely locating where the dissipation occurs, makes the process of acquiring information (and in particular its energy cost) something that is thought to be intrinsic to the system. Something that is thought to be independent of us.
Let us see how Landauer achieved this result.

\subsection{Landauer's derivation of his principle}\label{Landauer_derivation}

Landauer's derivation lies in four items:

\begin{enumerate}
	\item The concrete implementation of a bit of memory is necessarily achieved via a physical system which must exhibit two stable thermodynamic states of equilibrium denoted 0 and 1, e.g. a particle in a bistable potential or a particle in a box divided in two.
	
	\item The procedure for erasing (the erasure) a bit of memory, say setting the bit-value to state 0, must be independent of its initial value (0 or 1), because it must be able to work for an unknown value.
	
	\item The two different paths that take the bit from state 0 or from state 1 to state 0 necessarily merge into one, at a point (called state S) where the value of the bit is undetermined. Before this point, both paths are judged necessarily uncontrolled and thermodynamically irreversible, because in the reverse direction, state S is a bifurcation, which is assumed not to be tractable for a controlled deterministic procedure. The consequence is that neither work nor heat can be obtained from the system during this step between state 0 or 1 and state S. 
	
	\item According to Clausius inequality (Eq.\,\ref{Clausius_ineq1} or \ref{Clausius_ineq2}), the subsequent step, from state S to state 0, necessarily requires at least $T\ln2$ of work that will be dissipated as heat in the surroundings.
\end{enumerate}

The first step (item 3) is similar to the free expansion of a gas while the second (item 4) resembles an isothermal compression. The two together leads to the energy balance:
\begin{equation}
	W_{\text{erase/bit}}\ge T\ln 2
\end{equation}
This inequality is known as Landauer bound and constitute the core of Landauer principle.

\begin{figure}[!htbp]
	\begin{center}
		\includegraphics[width=0.6\linewidth]{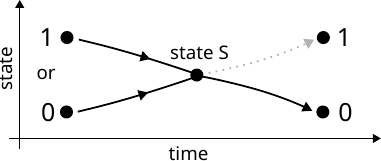}
		\caption{Landauer erasure: according to Landauer, the junction of two paths at a point called state S makes the process before this point necessarily uncontrolled and thermodynamically irreversible (just like the free expansion of gas).}
		\label{fifurc}
	\end{center}
\end{figure}

Landauer principle and Brillouin negentropy principle of information (section \ref{reformulation_Clausius}) are often confused. This point should be clarified.

\subsection{Landauer erasure vs. Brillouin negentropy}\label{compare1}

Let us bring 
Brillouin and Landauer inequalities together:
\begin{eqnarray}
	\text{Brillouin:}&\quad 	W_\text{acq/bit} &\ge T \ln 2 \\
	\text{Landauer:} &\quad 	W_{\text{erase/bit}}&\ge T\ln 2
\end{eqnarray}
and see how they differ and resemble.

A first major difference between the two principles which is not immediately apparent concerns their respective robustness.
Brillouin negentropy principle comes\,: 1)~from the 2nd law of thermodynamics and from the Clausius inequality, which have been confirmed by all known experiments; and 2)~from two mathematical results: Gibbs's and Shannon's ones. 
Since mathematical results are demonstrations that are not affected by the problem of induction posed in the natural sciences, it follows that the validity of the principle of negentropy is ultimately only conditioned by the validity of the 2nd law.
Brillouin negentropy principle will remain true as long as the 2nd law does.
This makes us very confident.
As for Landauer principle, it also uses the Clausius inequality (item~4, \S\ref{Landauer_derivation}), but it is also conditioned by the validity of the three preceding hypothetic items. Landauer considered a special case, without any assurance that it could be generalized. This makes us less confident.

Brillouin negentropy principle can be considered a fundamental general principle of physics. The status of Landauer principle is more uncertain.
This point cannot be denied and that is why the two principles should not be confused. 

Brillouin principle concerns the acquisition of one bit of information (so encoding the source outcomes requires one bit less). Landauer principle concerns the erasure of a data-bit. An in-depth comparison is done in the next section \S\ref{invalidation}, but here let us just deal with the following point. Although the meaning of the two inequalities is very different, both lead to the same result in the case where the acquisition of information consists of a cyclic sequence of operations involving erasure. As for example the cyclic operations performed by a Szilard demon having a finite size of memory. In this case both inequalities give the energy cost that must be paid by the demon for each cycle. So that the energy produced by the device does not come from nothing. Although conceptually different, the two solutions to the problem posed by the demon are equivalent.	
	The only difference is that Brillouin inequality concerns the net balance of an overall acquisition cycle (that may include erasure); whereas Landauer inequality stipulates where exactly in the cycle the energy cost is paid (in the erasure).
	The only difference is a question of (non-)localization.
	
According to the last point, it is not possible to consider that Landauer erasure principle solves problems posed by demons that would not be solved by Brillouin negentropy principle, unless one considers that the non-localization of the dissipation poses problem. Let us report some recent quotes among many others on this matter:

\myquote{The Landauer principle is one of the cornerstone of the modern theory of information} (Herrera\,\cite{Herrera2020})

\myquote{Landauer’s principle was central to solving the
paradox of Maxwell’s demon.} (Lutz \& Ciliberto\,\cite{Lutz_2015})

\myquote{Informational theory is usually supplied in a form that is independent of any physical embodiment. In contrast, Rolf Landauer in his papers argued that information is physical and it has an energy
equivalent.} (Bormashenko\,\cite{Bormashenko_2019b}). In other words, Landauer principle is more physical than Brillouin's.

\myquote{Since [Szilard's work] in 1929, the story has remained much the same. The	only major change was made by Landauer, who suggested that the erasure of information was specifically what generated heat.} (Witkowski et al.\,\cite{Witkowski2024}). As a reminder, Shannon's work dates from 1948, Brillouin's from 1953, Landauer's from 1961.

Landauer's solution is clearly preferred and the non-localization of dissipation therefore does indeed pose a problem.

Acquisition of information cycles are cyclic sequences of operations, they are not periodic sequences of thermodynamic states since the incoming value of data varies from cycle to cycle.
An acquisition cycle is actually a process that takes the acquisition device from state A to state B (different or not from A).
Let us compare with mechanical processes. Consider a body that is moved from point $A$ to point $B$. 
The minimum mechanical work to be provided for this is given by the difference of potential energy of the body between these two points, regardless of the path taken between the two.
There is no law of mechanics that dictates that work must be done precisely somewhere. There is no law of physics that tells us where work is done along the path.
Despite this, potential energy is considered to be physical.
The negentropy principle and how it solves the problem of demons operate exactly the same way.

Why something that does not pose problem with mechanics and energy, poses problem with thermodynamics and information\,?
Why is something, which is not necessary for energy to be considered physical, requested for information\,?

The reason is that with energy we can easily forget that it is a pure abstraction. We can easily forget that energy is not \textquote{materialized} by something like \textquote{an energy particle} (an object) that could exist independently of us (see \S\ref{first_principle}).
We can easily forget that energy is not something that belongs to the Platonic \textit{objective reality}.
With information we cannot. Localizing and materializing information is viewed as the solution.

The last attempt at localizing and materializing information is the principle of information-mass equivalence\,\cite{HERRERA2014, Vopson_2019, Vopson_2020, Herrera2020, Masic_2021, Vopson_2022}. In short: any piece of information is stored under the form of data-bits, these are supposed to store energy when set to a particular value, energy has a mass equivalence, thus information has a mass. Information becomes definitely physical, understand tangible and materialized.

Landauer's motivations in stating his principle were probably not of this nature, but in my opinion, the reasons of its success certainly are. Because there is no point in Landauer's reasoning that resists close examination.

\section{Invalidation of Landauer erasure principle}\label{invalidation}

Literature on information and Landauer's principle reveals a great deal of confusion in the vocabulary, such that we no longer know what we are talking about.
The confusion is about information and data-bit, known and unknown, logically reversible versus irreversible etc.
It is therefore imperative to clarify and untangle all these notions.

\subsection{Piece of information versus data-bit}\label{info_vs_bit}
	
 To the word information we give the usual meaning that is the primary sense of the Oxford Dictionary: \myquote{Information: The imparting of knowledge in general. Knowledge communicated concerning some particular fact, subject, or event}.
 
 This usual meaning is the one retained in the literature on this subject:
 \myquote{But what is information? A simple, intuitive answer is “what you don’t already know”}\,\cite{Lutz_2015}. \myquote{If someone tells you that the earth is spherical, you surely would not learn much}\,\cite{Ciliberto_2018}.
 
 This meaning is also Shannon's (see end of \S\ref{Quantity_info}), who additionally made information quantifiable. So that, acquiring one bit of information about a dynamical source means that one bit less is required for the encoding of the outcomes, the uncertainty we have in describing the system decreases.
 But information cannot be given twice. Acquiring a second time the same piece of information about a source does not save any more encoding space.
 
 Any piece of information is necessarily stored in memory on a material support composed of material units called data-bits for convenience. A data-bit is a material object which displays two distinguishable stationary configurations (0 and 1).
 A data-bit can be duplicated, or copied, and this increases the amount of memory used but leaves the quantity of information unchanged.
 In other words, one bit of information is necessarily stored under the form of a data-bit set to a particular value. But conversely, a data-bit set to a particular value is not always information. Because, many copies of a data-bit contain no more information than the original alone.
 	
A data-bit is necessarily located somewhere, while a piece of information is not. Consider a newly acquired piece of information stored under the form of a given data-bit. Make a copy of it, which is by definition located elsewhere.
Regarding the information, the original does not have a special status compared to the copy. The original and the copy are two distinct physical objects but do not encode for two distinct pieces of information.
Erase the original or the copy: the information is still there. Both must be erased so that the information is irretrievably destroyed.
If there exists many copies of an original data-bit storing a piece of information, the corresponding information is not \textquote{diluted} or shared among each copy. 
Duplicating or erasing one copy or the original, leaves the amount of information unchanged, provided one copy still exists. 
Neither the original data-bit nor the copies contain a small portion of information that could therefore be considered to be located on that data-bit. The same is true if an original data-bit is unique.

In terms of computer science, a data-bit is identified by a pointer whereas a piece of information by a value. 

It is true that any piece of information needs a material support in the form of data-bits, but this material support does not make the piece of information localized on this support. 
Although information requires matter, it is non-material and remains intangible and abstract.
This may seem paradoxical, but actually other abstract physical concepts require matter but are never materialized: energy, fields, time, interactions (we are still looking for the graviton)...
Information is physical and needs a material embodiment,
but that does not mean it must be confused with its physical embodiment, that does not mean it must be confused with its physical support. Both must be clearly dissociated. 
This is a first misunderstanding to avoid.

Landauer contributes to the confusion, for instance:
\myquote{Information is not a disembodied abstract entity; it is always tied to a physical representation. It is represented by engraving on a stone tablet, a spin, a charge, a hole in a punched card, a mark on paper, or some other equivalent}\,\cite{Landauer1996}.
The fact that information can be embodied in multiple ways proves precisely that it is abstract and cannot be confused with its embodiment: embodiment changes but information does not. Moreover, the fact that information must be embodied also proves that it is abstract: matter does not need to be embodied, nor does a data-bit.

\subsection{(ir)recoverable loss of information versus thermodynamic (ir)reversibility}\label{irrev1}

Another way to think about the crucial difference between abstract information and material data-bit is that a data-bit is a thermodynamic system in itself, whereas a piece of information is not.

In particular, a data-bit can undergo thermodynamic processes such as erasure. That is to say, processes that only involve the data-bit (the thermodynamic system) and the part of the surroundings which interacts with it. Given an initial state and a final state of the data-bit, whether or not the process is thermodynamically reversible depends solely on the heat exchanged with the surroundings via the Clausius inequality (Eq.\,\ref{Clausius_ineq1}). Given these two states of the data-bit, the thermodynamic (ir)reversibility of a process is a property of the path.

On the contrary, the irreparable destruction of a piece of information, is not only conditional on the erasure of one given local data-bit, but is also conditional on the non-existence of a copy somewhere in the global system. Whether or not a local process causes a global loss of information depends on the initial state and final state of the global system. The irreparable destruction of a piece of information that a process can cause is a property of these two states. It is not a property of the path\,\cite{Lairez_2024b}.

In other words, we can erase a given data-bit in a thermodynamically reversible or irreversible way, but at the end, whatever the way, the initial and final states of the global system is the same. Likewise, we may, or may not, irretrievably lose a piece of information by erasing a given data-bit. But if an eventual copy is far away from the erased data-bit, the local erasing process does not \textquote{know} whether or not the copy exists. So that its thermodynamic (ir)reversibility is independent of the existence of the copy (unless we consider that everything interact with everything, but then, not only thermodynamics must be reconstructed, but all of physics).

The thermodynamic (ir)reversibility of a process undergone by a given local data-bit and the (ir)recoverable loss of information of the global system\,\cite{Lairez_2024b} are totally independent (to say otherwise is either a mistake or comes from an implicit different meaning given to the word information).
This is due to the mere fact that information is a non-material abstraction whereas a data-bit is not.

\subsection{Logical irreversibility:  equivocal operation or irrecoverable loss of information\,?}

This brings us to introduce and clarify the term \textquote{logical irreversibility} often used in relation with information and thermodynamic irreversibility.

Let us consider mathematical functions which apply on a finite subset of integers and can be entirely defined by a table consisting in two columns with input and output values. Bijections (one-to-one functions) allow the input value to be determined unequivocally from knowledge of the output. But, mathematical functions like $\textsc{sqr}:a \to a^2$ or $\textsc{abs}:a \to \abs{a}$  are not bijection, so that one given output value does not unequivocally define an input. The function $\textsc{zero}: a \to 0$ enters in this last category that will be called \textquote{equivocal} functions.

When considering a function like \textsc{erase} (or \textsc{set to 0}), implicitly, we are not considering an abstract mathematical function that applies to numbers having a given value. A number cannot be erased, nor set to a particular value. The number $\pi$ or the number 2 cannot be erased. When considering \textsc{erase}, we are considering the physical implementation of a mathematical equivocal function in a computer that applies on material entities such as data-bits. \textsc{erase} does not apply on a value but applies on the data-bit pointed by its argument.
Although the \textsc{erase} function is equivocal, whether or not it causes a loss of information depends on the existence of a copy of the erased data-bit (see previous section).

The meaning given by authors to the words \textquote{logical (ir)revesibility} is not clear, even if we restrict ourselves to the papers of Landauer and Bennett. 

\begin{enumerate}
	\item At first, \myquote{We shall call a device logically irreversible if the output of a device does not uniquely define the inputs}\,\cite{Landauer_1961}.
	\myquote{[\textsc{erase}] is an example of a logical truth function which we shall call irreversible}\,\cite{Landauer_1961}.
	
	Which means that \textquote{logically irreversible} functions applied to a particular data-bit are those that are equivocal, regardless of whether or not they cause an irrecoverable loss of information.
		
	\item But in the same paper: \myquote{At first sight it may seem that logical reversibility is simply obtained by saving the input in some corner of the machine. We shall, however, label a machine as being logically reversible, if and only if all its individual steps are logically reversible.}\,\cite{Landauer_1961}
	And some years later:
	\myquote{Computation that preserves information at every step along the way (and not just by trivially storing the initial data) is called reversible computation}\,\cite{Landauer_1991}. \myquote{It is easy to render any computer reversible in a rather trivial sense, by having it save all the information it would otherwise throw away}\,\cite{Bennett_1982}. 
	
	From which we can legitimately conclude that logical irreversibility is linked to the irrecoverable loss of information that can be avoided by making copies.
\end{enumerate}

We are a little lost.
In the sentence \myquote{Logical irreversibility, we believe, in turn implies physical [thermodynamical] irreversibility}\,\cite{Landauer_1961}, by logical irreversibility, does Landauer have in mind an equivocal operation on a data-bit or an irrecoverable loss of information\,?

For our purpose, the last option has already been addressed in \S\ref{irrev1}, the first remains. 

\subsection{Known versus unknown value of a data-bit}

The universe contains systems with computing capabilities and memory (human brains, computers, and even demons) which will be called computers for convenience.
Let us consider a data-bit and a computer. The value of the data-bit is considered as known by the computer if the data-bit or a copy is located inside the computer memory. Otherwise it is unknown.

The data bit is considered as a thermodynamic system on which the computer is supposed to be able to apply a process, regardless of its location, while being able to observe the process. The computer is therefore also the observer and judges the thermodynamic reversibility of the process according to the Clausius inequality 
and to his knowledge about the initial and final states of the data-bit.

Known or unknown initial value of the data-bit probably lead to different judgments about the thermodynamic reversibility of the data-bit erasure. But this has absolutely no incidence on the global quantity of information: If the data-bit has no copy, the erasure will result in a loss of global information, otherwise it will not.

\subsection{On the thermodynamic reversibility of Landauer erasure}

\begin{figure}[!htbp]
	\begin{center}
		\includegraphics[width=0.8\linewidth]{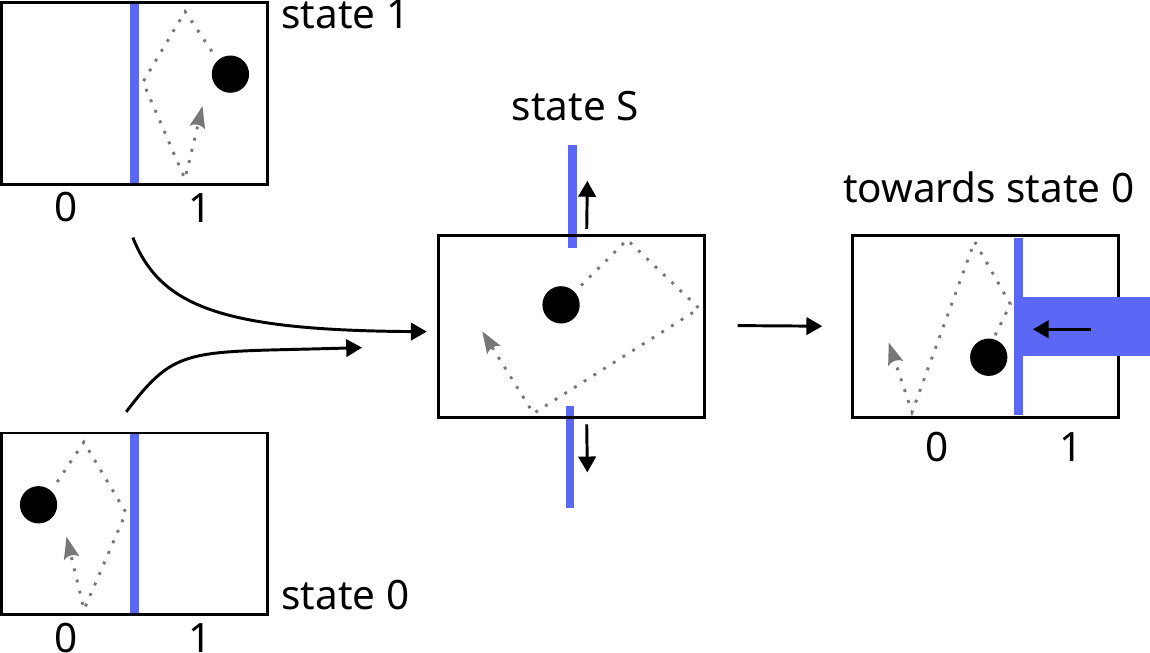}
		\caption{Landauer erasure of a data-bit represented by a single particle in a box divided in two compartments. By removing the partition whatever its initial position the particle freely expands to the state S. Then, a piston performs an isothermal compression until the particle is in state 0.}
		\label{Landauer_erasure}
	\end{center}
\end{figure}

Let us consider the physical implementation of a data-bit that has been chosen by Landauer with a representation directly inspired by the Szilard engine\,\cite{Plenio_2001, Maroney2005}. A single particle is in a box partitioned in two (see Fig.\,\ref{Landauer_erasure}). If the particle is in the left-hand side the bit-value is 0. If it is in the right-hand side the bit value is 1. When the partition is removed the data-bit is in state S. The erasure obeys the following sequence of steps:
\begin{enumerate}
	\item From state 0 or 1 to state S: remove the partition.
	\item From state S to state 0: perform an isothermal (thermodynamically reversible) compression with a piston.
\end{enumerate}

\begin{figure}[!htbp]
	\begin{center}
		\includegraphics[width=0.8\linewidth]{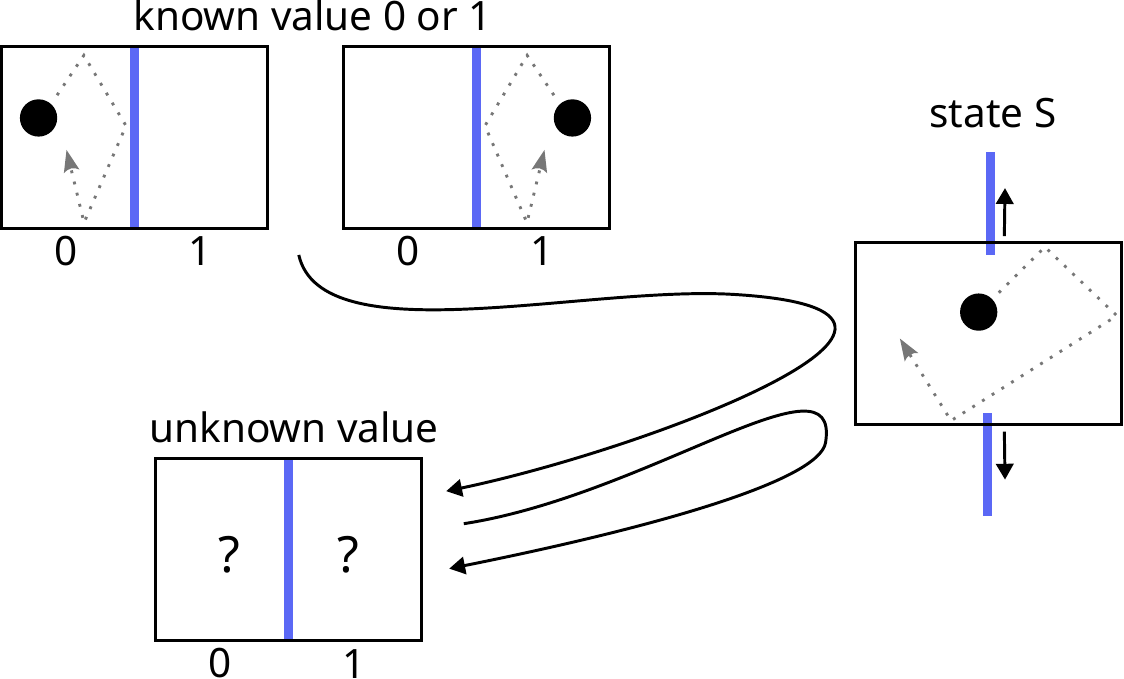}
		\caption{Landauer erasure of a data-bit: in case the initial value of the bit is unknown the first step of the procedure (see Fig.\,\ref{Landauer_erasure}) from state 0 or state 1 to state S is thermodynamically irreversible (by simply putting back the partition the system does not recover its initial state); in case the initial value is unknown the same process is thermodynamically reversible.}
		\label{known_unknown}
	\end{center}
\end{figure}

According to Landauer, the thermodynamic irreversibility of the erasure comes from step 1 that is identified as being free expansion of a single-particle \textquote{gas}. And this is the point, the free expansion of a gas, or the mixing of two gases, are thermodynamically irreversible or not depending on our knowledge about the initial state (see \S\ref{free_exp} and \S\ref{Gibbs_paradox1}). Here, this will depend on whether or not we know the initial value of the data-bit. Hence the two possibilities: 1)~for a known value of the data-bit the first step of Landauer erasure from state 0 or 1 to state S is thermodynamically irreversible; 2)~for unknown value it is reversible\,\cite{Maroney2005} (see Fig.\,\ref{known_unknown}).

Once this first step achieved, however the bit is not yet erased. The second step remains to be done from state S to state 0, that dissipates $T\ln2$ of heat. Question: on the one hand, the thermodynamic reversibility of global erasure is conditional on that of the first step; but on the other hand, how is it then that the dissipated heat is not zero for the reversible case of an unknown value\,? This is because the thermodynamic cycle is not yet closed. 
It remains to bring back the data-bit from state 0 to an unknown-state. In the framework of Landauer implementation this can be achieved via the following sequence\,:
\begin{enumerate}\setcounter{enumi}{2}
	\item From state 0 to state S: use the piston that is already in position to do an isothermal expansion (i.e. reverse step 2 of Landauer erasure).
	\item From state S to unknown-state: put back the partition, so that the final state is random and unknown.
\end{enumerate}
The energy exchanged during the step~3 compensates exactly that of step 2, so that the net balance is 0. Step~4 puts the data-bit at a random unknown value exactly as it was before. The Landauer erasure for an unknown value is thermodynamically reversible.

For an initially known value, the above step~4 must be replaced by\,:
\begin{enumerate}\setcounter{enumi}{3}
	\item From state S to the initial state (0 or 1): use a piston to push the particle into the proper compartment.
\end{enumerate}
This will cause $T\ln 2$ of heat dissipation, so that the Landauer procedure to erase the data-bit is actually irreversible for a known value and only in this case.

This difference between known and unknown value, which tarnishes the universal character of Landauer erasure principle, was initially completely ignored by Landauer\,\cite{Landauer_1961}, but finally acknowledged by Bennett\,\cite{Bennett_2003}: \myquote{If a logically irreversible operation like erasure is applied to random data, the operation still
may be thermodynamically reversible [...]. But if, as is more
usual in computing, the logically irreversible operation is applied to known data, the operation is thermodynamically irreversible}.
Surprisingly, this nuance brought by Bennett himself, however, seems to be completely ignored by those who take up Landauer erasure principle and claim it to be universal. While the case of a known value for a bit of data is probably \myquote{more usual in computing}\,\cite{Bennett_2003}, it seems to me impossible to assert the same thing in absolute terms for any data-bit in the universe as Landauer intends to do\,\cite{Landauer_1991}.

\subsection{On the thermodynamic reversibility of alternative erasures}

By following the Landauer procedure to erase a data-bit the process is thermodynamically irreversible in case its initial value is known. But this procedure is not a general case. Alternatives can be proposed allowing erasure to be quasi-static in all cases (known or unknown initial value).
Two possibilities are exposed in the following. The first questions the claim that thermodynamic states necessarily map to logical states (see \S\ref{Landauer_derivation} item~1) and is reminiscent to the incomplete information thermodynamics deals with, which has already been evoked for the Gibbs paradoxes (\S\ref{Gibbs_paradox1} and \S\ref{Gibbs_paradox2}). The second questions the idea that the path from state 0 or 1 to state S is necessarily uncontrolled and non-quasistatic (see \S\ref{Landauer_derivation} item~3) because it merges two ways so that in the other direction it displays a bifurcation.

\subsubsection{States incompletely defined for thermodynamics}

The Gibbs paradoxes (\S\ref{Gibbs_paradox1} and \S\ref{Gibbs_paradox2}) arise when we mistakenly assume that thermodynamics treats the states of a system as something intrinsic to the system, such as the total information needed to describe it entirely (if that is possible).
Actually thermodynamics does not, and whether or not two states differ in thermodynamics depends on the observer and his knowledge. Based on this, counter-examples to the generality of Landauer erasure have been proposed\,\cite{Lairez_2023, Lairez_2024a}. Basically, they work like that: the logical value of the data-bit is evaluated on the basis of a greatest quantity of information than that needed to describe the thermodynamic state. Typically, one single thermodynamic state can correspond to two distinct logical states. Then, there is absolutely no impediment for the erasure to be quasistatic.

\subsubsection{Managing two possibilities with a single externally imposed procedure}

With the Landauer erasure procedure, the externally controlled process to drive the particle in an undetermined state (state S) and from there to state 0, is conceived as an external force applied to the particle. Therefore, according to Newton's deterministic mechanics, this is not possible because a unique time-varying force cannot cause two trajectories. The process in question is thus necessarily thermodynamically irreversible.

\begin{figure}[!htbp]
	\begin{center}
		\includegraphics[width=1\linewidth]{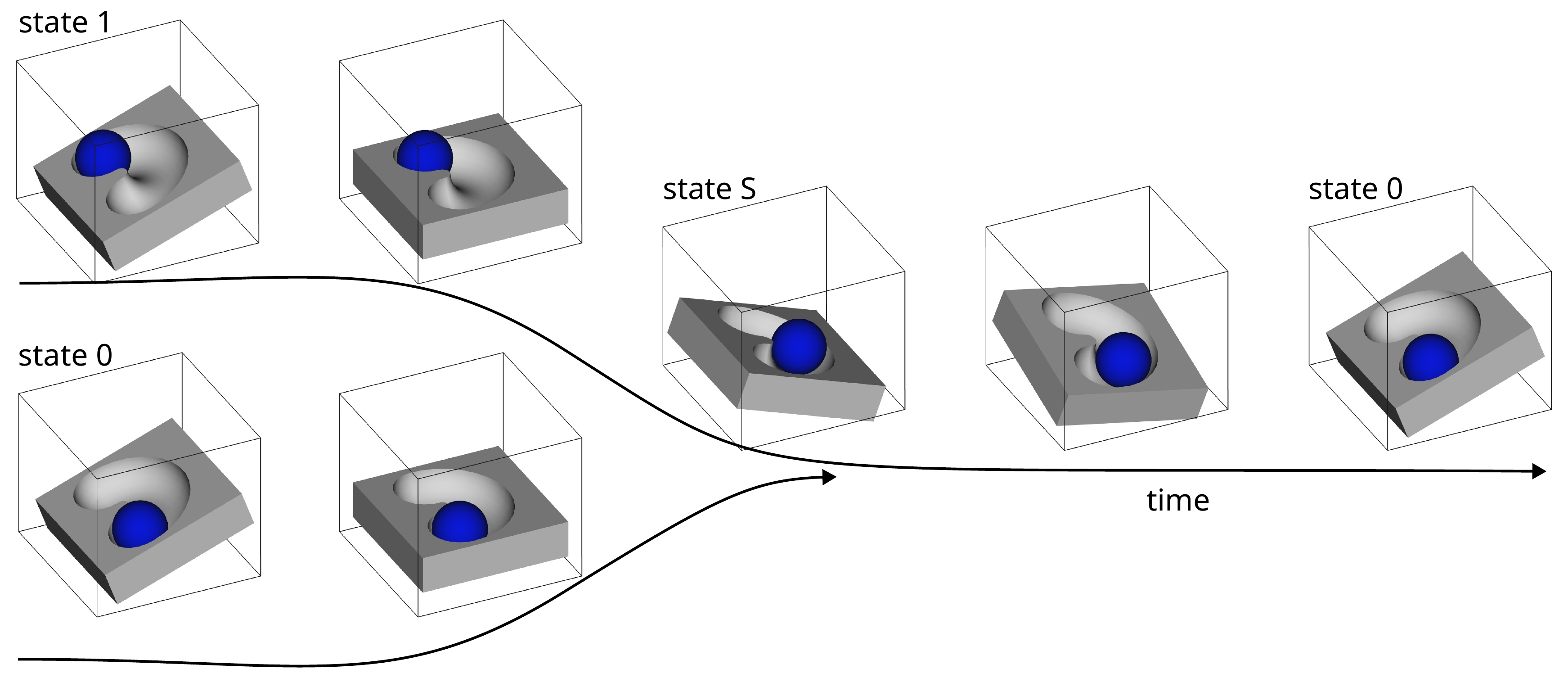}
		\caption{Data-bit materialized by a particle in a topographic relief (a short range repulsive surface) plus a vertical attractive field. Regardless of the initial bit-value, the same time-varying tilt of the surface relative to the field allows for quasistatic erasure.}
		\label{field}
	\end{center}
\end{figure}

However, an externally controlled process may not necessarily be a force directly applied to the particle.
It can be an externally applied time-varying field, that exerts on a particle in a fixed and constant short range potential landscape, which can be viewed as a topographic relief\,\cite{Lairez_2024b}. 
The topographic relief is part of the physical implementation of the data-bit and manages the two possible trajectories. 
The force experienced by the particle results from these two contributions: one external (the time-varying field), the other internal (the topographic relief).
So that a single externally controlled process (the time-varying field) can result in two different trajectories depending on the initial position of the particle. There is no longer impediment for such an externally controlled process be thermodynamically quasistatic. 
Figure \ref{field} gives an example of implementation. The motion of the particle follows the direction of maximum potential descent and can be as slow as desired.
\subsection{Supposed information-mass equivalence}

The principle of information-mass equivalence is presented as a consequence of Landauer erasure principle\,\cite{HERRERA2014, Vopson_2019, Vopson_2020, Herrera2020, Masic_2021, Vopson_2022}. So that, by invalidating the latter, we invalidate the former. 
But in reality, even if we accept the generality of Landauer erasure principle (ignoring, for example, the counterexamples given in the previous section), the information-mass equivalence principle is above all the natural consequence of several erroneous ideas\,\cite{Lairez_2024a}.

\begin{enumerate}
	\item The idea that data-bits necessarily store information.
	
	 In reality, information is necessarily stored under the form of data-bits but data-bits (even those set to a particular value according to a given acquired piece of information) do not necessarily store information. Copies do not. Information is not materialized by data-bits. This point has been developed extensively in \S\ref{info_vs_bit}.
	
	\item The idea that Brillouin negentropy principle and Landauer erasure principle are the same (see \S\ref{compare1}) or that the two can be amalgamated: 
	\myquote{The mass of a bit of information and the Brillouin's principle}\,\cite{HERRERA2014} in 2014 becomes 6 years later: \myquote{We shall first analyze, in	the context of general relativity, the consequences derived from the fact, implied by Landauer principle, that information has mass}\,\cite{Herrera2020}
	
In reality, the Brillouin principle of negentropy tells us that a minimum amount of work is needed to reduce the uncertainty we have about the system, that is, to acquire one bit of information.

Whereas Landauer erasure principle, for its part, claims that a data-bit must receive a minimum of energy to be erased. 

It follows that the two are only compatible in the context of acquiring information about a dynamical system, an acquisition which is carried out cyclically and which involves the Landauer erasure of the data-bits which were used to store outdated information, i.e. those of the previous cycles. Outside this context, the two are incompatible. We currently have no indication that the dynamic behavior of the Universe is cyclic, but if it were, one thing is certain: all the information we have was acquired during the same cycle.
	 
	\item The idea that information cannot be outdated (then it would no longer be information) so that a data-bit stores information forever, which is closely related to the idea that a piece of information is a self-sufficient entity and that there is no need to specify information about what:
	\myquote{To test the hypothesis [the mass-energy-information equivalence principle] we propose here an experiment, predicting that the mass of a data storage device would increase by a small amount when is full of digital information relative to its mass in erased state. For 1Tb device the estimated mass change is $2.5 \times 10^{-25}$\,Kg.} (Vopson\,\cite{Vopson_2019}). 

	In reality, information about old cycles of a Szilard engine is of no use. It is outdated. It is for this reason that the demon can erase the data-bit used for its storage.
	
	Let us imagine a Szilard demon who would store all the information about each cycles until its hard-disk is full. According to the Brillouin negentropy principle, the cost of acquiring information is paid cycle-by-cycle, so that the net energy balance between work produced and the cost of information is always zero at the end of each cycle. According to Landauer's erasure principle the information cost will be never paid until the hard-disk is erased. But here there is a real inconsistency: does the filled hard drive contain more energy and have a greater mass (according to a misunderstanding of the Brillouin negentropy principle) or does it need energy to be erased, so that in the end once erased the mass is greater (according to Landauer erasure)? In fact, neither.
	
	\item The idea that the work required to acquire information is stored as energy in the data-bit, just as mechanical work exerted on a body can be stored within it as potential energy, but actually under the form of rest mass\,\cite{Brillouin_1965a, Brillouin_1965b}.
	
	In reality, at constant temperature, when a system received work that decreases its entropy (that reduces the uncertainty we have, or that reduces the information we lack about it), this work is not stored as internal energy, it is dissipated as heat in the surroundings.
	A volume of gas that can be compressed or expanded isothermally with a piston behaves in exactly the same way as if it were a spring. But the latter stores potential energy (thus increases its rest mass), while the former does not. Actually for the spring, the potential energy at the macroscopic scale originates from the interaction potential between particles. In a perfect gas, there is no interaction potential, except that of colliding hard spheres. The force exerted on the piston is an emergent property.
\end{enumerate}

Once these points are clarified, the information no longer has a mass equivalent. It remains an abstraction.

\section*{Conclusion}

The role of the observer in physics is always a disturbing point. Despite this, it is recognized in quantum mechanics\,\cite{Rosenblum2002}: particles have no identity until they reach a detector; Schr\"odinger's cat is both dead and alive until we observe its state. This role is also recognized in special\,\cite{Feynman_1963_relativity} and general\,\cite{Bini2014} relativity, where everything is relative to the observer.
In these areas, the role of the observer is problematic, but it is accepted. Probably because quantum and relativity theories are already so abstract, the subject matter so far removed from our scale, that this role is just another abstraction, a strangeness added to so many others.
With thermodynamics, we are initially in the concrete realm of engines, of machines, everything is on our scale. So the role of the observer is probably more unexpected, but on reflection not so surprising.

\myquote{Any serious consideration of a physical theory must take into account the distinction between the objective reality, which is independent of any theory, and the physical concepts with which the theory operates. These concepts are intended to correspond with the objective reality, and by means of this concepts we picture this reality to ourselves} (Einstein et al.\,\cite{Einstein1935}).
Energy is one of such concepts, it was not discovered by Joule, it was invented.
This invention is one of the most useful for describing the physical world, but it remains an abstraction nonetheless.
The concept of entropy and that of information are in the same vein.
Yes, information is physical. Just like energy is, but no more.

\bibliography{dicho.bib}

\begin{thebibliography}{96}%
\makeatletter
\providecommand \@ifxundefined [1]{%
 \@ifx{#1\undefined}
}%
\providecommand \@ifnum [1]{%
 \ifnum #1\expandafter \@firstoftwo
 \else \expandafter \@secondoftwo
 \fi
}%
\providecommand \@ifx [1]{%
 \ifx #1\expandafter \@firstoftwo
 \else \expandafter \@secondoftwo
 \fi
}%
\providecommand \natexlab [1]{#1}%
\providecommand \enquote  [1]{``#1''}%
\providecommand \bibnamefont  [1]{#1}%
\providecommand \bibfnamefont [1]{#1}%
\providecommand \citenamefont [1]{#1}%
\providecommand \href@noop [0]{\@secondoftwo}%
\providecommand \href [0]{\begingroup \@sanitize@url \@href}%
\providecommand \@href[1]{\@@startlink{#1}\@@href}%
\providecommand \@@href[1]{\endgroup#1\@@endlink}%
\providecommand \@sanitize@url [0]{\catcode `\\12\catcode `\$12\catcode
  `\&12\catcode `\#12\catcode `\^12\catcode `\_12\catcode `\%12\relax}%
\providecommand \@@startlink[1]{}%
\providecommand \@@endlink[0]{}%
\providecommand \url  [0]{\begingroup\@sanitize@url \@url }%
\providecommand \@url [1]{\endgroup\@href {#1}{\urlprefix }}%
\providecommand \urlprefix  [0]{URL }%
\providecommand \Eprint [0]{\href }%
\providecommand \doibase [0]{https://doi.org/}%
\providecommand \selectlanguage [0]{\@gobble}%
\providecommand \bibinfo  [0]{\@secondoftwo}%
\providecommand \bibfield  [0]{\@secondoftwo}%
\providecommand \translation [1]{[#1]}%
\providecommand \BibitemOpen [0]{}%
\providecommand \bibitemStop [0]{}%
\providecommand \bibitemNoStop [0]{.\EOS\space}%
\providecommand \EOS [0]{\spacefactor3000\relax}%
\providecommand \BibitemShut  [1]{\csname bibitem#1\endcsname}%
\let\auto@bib@innerbib\@empty
\bibitem [{\citenamefont {Descartes}(2008)}]{Descartes2008}%
  \BibitemOpen
  \bibfield  {author} {\bibinfo {author} {\bibfnamefont {R.}~\bibnamefont
  {Descartes}},\ }\href@noop {} {\emph {\bibinfo {title} {Meditations on First
  Philosophy}}},\ edited by\ \bibinfo {editor} {\bibfnamefont {M.}~\bibnamefont
  {Moriarty}},\ Oxford World's Classics Ser.\ (\bibinfo  {publisher} {Oxford
  University Press USA - OSO},\ \bibinfo {address} {Oxford},\ \bibinfo {year}
  {2008})\BibitemShut {NoStop}%
\bibitem [{\citenamefont {Carter}(1931)}]{Carter_1931}%
  \BibitemOpen
  \bibfield  {author} {\bibinfo {author} {\bibfnamefont {R.}~\bibnamefont
  {Carter}},\ }\href@noop {} {\emph {\bibinfo {title} {Descartes' medical
  philosophy : the organic solution to the mind-body problem}}}\ (\bibinfo
  {publisher} {The Johns Hopkins University},\ \bibinfo {year}
  {1931})\BibitemShut {NoStop}%
\bibitem [{\citenamefont {Mach}(1919)}]{Mach_1919}%
  \BibitemOpen
  \bibfield  {author} {\bibinfo {author} {\bibfnamefont {E.}~\bibnamefont
  {Mach}},\ }\href@noop {} {\emph {\bibinfo {title} {The Science of
  Mechanics}}}\ (\bibinfo  {publisher} {The Open Court Publishing Company},\
  \bibinfo {year} {1919})\BibitemShut {NoStop}%
\bibitem [{\citenamefont {Duhem}(2021)}]{Duhem_2021}%
  \BibitemOpen
  \bibfield  {author} {\bibinfo {author} {\bibfnamefont {P.}~\bibnamefont
  {Duhem}},\ }\href {https://doi.org/10.2307/j.ctv1nj34vm} {\emph {\bibinfo
  {title} {The Aim and Structure of Physical Theory}}}\ (\bibinfo  {publisher}
  {Princeton University Press},\ \bibinfo {year} {2021})\BibitemShut {NoStop}%
\bibitem [{\citenamefont {Einstein}(1934)}]{Einstein_1934}%
  \BibitemOpen
  \bibfield  {author} {\bibinfo {author} {\bibfnamefont {A.}~\bibnamefont
  {Einstein}},\ }\bibfield  {title} {\bibinfo {title} {On the method of
  theoretical physics},\ }\href {https://doi.org/10.1086/286316} {\bibfield
  {journal} {\bibinfo  {journal} {Philosophy of Science}\ }\textbf {\bibinfo
  {volume} {1}},\ \bibinfo {pages} {163} (\bibinfo {year} {1934})}\BibitemShut
  {NoStop}%
\bibitem [{\citenamefont {Einstein}\ \emph {et~al.}(1935)\citenamefont
  {Einstein}, \citenamefont {Podolsky},\ and\ \citenamefont
  {Rosen}}]{Einstein1935}%
  \BibitemOpen
  \bibfield  {author} {\bibinfo {author} {\bibfnamefont {A.}~\bibnamefont
  {Einstein}}, \bibinfo {author} {\bibfnamefont {B.}~\bibnamefont {Podolsky}},\
  and\ \bibinfo {author} {\bibfnamefont {N.}~\bibnamefont {Rosen}},\ }\bibfield
   {title} {\bibinfo {title} {Can quantum-mechanical description of physical
  reality be considered complete?},\ }\href
  {https://doi.org/10.1103/physrev.47.777} {\bibfield  {journal} {\bibinfo
  {journal} {Physical Review}\ }\textbf {\bibinfo {volume} {47}},\ \bibinfo
  {pages} {777} (\bibinfo {year} {1935})}\BibitemShut {NoStop}%
\bibitem [{\citenamefont {Maxwell}(1872)}]{Maxwell_1872}%
  \BibitemOpen
  \bibfield  {author} {\bibinfo {author} {\bibfnamefont {J.~C.}\ \bibnamefont
  {Maxwell}},\ }\href
  {https://books.google.fr/books?id=5u84AAAAMAAJ&printsec=frontcover&hl=fr#v=onepage&q&f=false}
  {\emph {\bibinfo {title} {Theory of heat}}},\ \bibinfo {edition} {3rd}\ ed.\
  (\bibinfo  {publisher} {Longmans, Green and Co.},\ \bibinfo {address}
  {London},\ \bibinfo {year} {1872})\BibitemShut {NoStop}%
\bibitem [{\citenamefont {Planck}(1903)}]{Planck_1903}%
  \BibitemOpen
  \bibfield  {author} {\bibinfo {author} {\bibfnamefont {M.}~\bibnamefont
  {Planck}},\ }\href {https://www.gutenberg.org/files/50880/50880-pdf.pdf}
  {\emph {\bibinfo {title} {Treatise of thermodynamics}}}\ (\bibinfo
  {publisher} {Longmans, Green and Co.},\ \bibinfo {year} {1903})\BibitemShut
  {NoStop}%
\bibitem [{\citenamefont {Maxwell}(1878)}]{Maxwell_1878}%
  \BibitemOpen
  \bibfield  {author} {\bibinfo {author} {\bibfnamefont {J.~C.}\ \bibnamefont
  {Maxwell}},\ }\bibfield  {title} {\bibinfo {title} {Diffusion},\ }\href
  {https://doi.org/10.1017/CBO9780511710377.064} {\bibfield  {journal}
  {\bibinfo  {journal} {Encyclopedia Britannica, reproduced in Scientific
  papers}\ }\textbf {\bibinfo {volume} {2}},\ \bibinfo {pages} {625} (\bibinfo
  {year} {1878})}\BibitemShut {NoStop}%
\bibitem [{\citenamefont {Joule}(1850)}]{Joule_1850}%
  \BibitemOpen
  \bibfield  {author} {\bibinfo {author} {\bibfnamefont {J.}~\bibnamefont
  {Joule}},\ }\bibfield  {title} {\bibinfo {title} {On the mechanical
  equivalent of heat},\ }\href {https://www.jstor.org/stable/108427} {\bibfield
   {journal} {\bibinfo  {journal} {Philosophical Transactions of the Royal
  Society of London}\ }\textbf {\bibinfo {volume} {140}},\ \bibinfo {pages}
  {61} (\bibinfo {year} {1850})}\BibitemShut {NoStop}%
\bibitem [{\citenamefont {Feynman}\ \emph {et~al.}(1966)\citenamefont
  {Feynman}, \citenamefont {Leighton},\ and\ \citenamefont
  {Sands}}]{Feynmann_Energy}%
  \BibitemOpen
  \bibfield  {author} {\bibinfo {author} {\bibfnamefont {R.~P.}\ \bibnamefont
  {Feynman}}, \bibinfo {author} {\bibfnamefont {R.~B.}\ \bibnamefont
  {Leighton}},\ and\ \bibinfo {author} {\bibfnamefont {M.}~\bibnamefont
  {Sands}},\ }\href {https://www.feynmanlectures.caltech.edu/I_04.html} {\emph
  {\bibinfo {title} {The {F}eynman lectures on physics}}}\ (\bibinfo
  {publisher} {Addison-Wesley, Reading, MA},\ \bibinfo {year} {1966})\
  Chap.~\bibinfo {chapter} {4}\BibitemShut {NoStop}%
\bibitem [{\citenamefont {Hecht}(2019)}]{Hecht_2019b}%
  \BibitemOpen
  \bibfield  {author} {\bibinfo {author} {\bibfnamefont {E.}~\bibnamefont
  {Hecht}},\ }\bibfield  {title} {\bibinfo {title} {Understanding energy as a
  subtle concept: A model for teaching and learning energy},\ }\href
  {https://doi.org/10.1119/1.5109863} {\bibfield  {journal} {\bibinfo
  {journal} {American Journal of Physics}\ }\textbf {\bibinfo {volume} {87}},\
  \bibinfo {pages} {495} (\bibinfo {year} {2019})}\BibitemShut {NoStop}%
\bibitem [{\citenamefont {Clausius}(1879)}]{Clausius_1879}%
  \BibitemOpen
  \bibfield  {author} {\bibinfo {author} {\bibfnamefont {R.}~\bibnamefont
  {Clausius}},\ }\href
  {https://books.google.fr/books?id=8LIEAAAAYAAJ&printsec=frontcover&hl=fr&source=gbs_ge_summary_r&cad=0#v=onepage&q&f=false}
  {\emph {\bibinfo {title} {The mechanical theory of heat}}}\ (\bibinfo
  {publisher} {Macmillan \& Co, London, UK},\ \bibinfo {year}
  {1879})\BibitemShut {NoStop}%
\bibitem [{\citenamefont {Jaynes}(1992)}]{Jaynes1992}%
  \BibitemOpen
  \bibfield  {author} {\bibinfo {author} {\bibfnamefont {E.~T.}\ \bibnamefont
  {Jaynes}},\ }\bibfield  {title} {\bibinfo {title} {The {G}ibbs paradox},\
  }in\ \href {https://doi.org/10.1007/978-94-017-2219-3_1} {\emph {\bibinfo
  {booktitle} {Maximum Entropy and Bayesian Methods}}}\ (\bibinfo  {publisher}
  {Springer Netherlands},\ \bibinfo {year} {1992})\ pp.\ \bibinfo {pages}
  {1--21}\BibitemShut {NoStop}%
\bibitem [{\citenamefont {Quine}(1976)}]{Quine1976}%
  \BibitemOpen
  \bibfield  {author} {\bibinfo {author} {\bibfnamefont {W.~V.}\ \bibnamefont
  {Quine}},\ }\href {https://archive.org/details/waysofparadox00quin/mode/2up}
  {\emph {\bibinfo {title} {The ways of paradox, and other essays}}}\ (\bibinfo
   {publisher} {Harvard University Press},\ \bibinfo {address} {Cambridge,
  Massachusetts},\ \bibinfo {year} {1976})\BibitemShut {NoStop}%
\bibitem [{\citenamefont {Gibbs}(1874)}]{Gibbs1874}%
  \BibitemOpen
  \bibfield  {author} {\bibinfo {author} {\bibfnamefont {J.~W.}\ \bibnamefont
  {Gibbs}},\ }\href {https://doi.org/10.5479/sil.421748.39088007099781} {\emph
  {\bibinfo {title} {On the equilibrium of heterogeneous substances : first
  [-second] part}}}\ (\bibinfo  {publisher} {Connecticut academy of arts and
  sciences},\ \bibinfo {year} {1874})\BibitemShut {NoStop}%
\bibitem [{\citenamefont {Lairez}(2024{\natexlab{a}})}]{Lairez_2024}%
  \BibitemOpen
  \bibfield  {author} {\bibinfo {author} {\bibfnamefont {D.}~\bibnamefont
  {Lairez}},\ }\bibfield  {title} {\bibinfo {title} {Thermostatistics,
  information, subjectivity: Why is this association so disturbing?},\ }\href
  {https://doi.org/10.3390/math12101498} {\bibfield  {journal} {\bibinfo
  {journal} {Mathematics}\ }\textbf {\bibinfo {volume} {12}},\ \bibinfo {pages}
  {1498} (\bibinfo {year} {2024}{\natexlab{a}})}\BibitemShut {NoStop}%
\bibitem [{\citenamefont {van Kampen}(1984)}]{van_Kampen_1984}%
  \BibitemOpen
  \bibfield  {author} {\bibinfo {author} {\bibfnamefont {N.}~\bibnamefont {van
  Kampen}},\ }\bibfield  {title} {\bibinfo {title} {The {G}ibbs paradox},\ }in\
  \href {https://doi.org/10.1016/b978-0-08-026523-0.50020-5} {\emph {\bibinfo
  {booktitle} {Essays in Theoretical Physics}}}\ (\bibinfo  {publisher}
  {Elsevier},\ \bibinfo {year} {1984})\ pp.\ \bibinfo {pages}
  {303--312}\BibitemShut {NoStop}%
\bibitem [{\citenamefont {Szilard}(1964)}]{Szilard_1964}%
  \BibitemOpen
  \bibfield  {author} {\bibinfo {author} {\bibfnamefont {L.}~\bibnamefont
  {Szilard}},\ }\bibfield  {title} {\bibinfo {title} {On the decrease of
  entropy in a thermodynamic system by the intervention of intelligent
  beings},\ }\href {https://doi.org/10.1002/bs.3830090402} {\bibfield
  {journal} {\bibinfo  {journal} {Behavioral Science}\ }\textbf {\bibinfo
  {volume} {9}},\ \bibinfo {pages} {301} (\bibinfo {year} {1964})}\BibitemShut
  {NoStop}%
\bibitem [{\citenamefont {Darrigol}(2021)}]{Darrigol_2021}%
  \BibitemOpen
  \bibfield  {author} {\bibinfo {author} {\bibfnamefont {O.}~\bibnamefont
  {Darrigol}},\ }\bibfield  {title} {\bibinfo {title} {{B}oltzmann's reply to
  the {L}oschmidt paradox: a commented translation},\ }\bibfield  {journal}
  {\bibinfo  {journal} {The European Physical Journal H}\ }\textbf {\bibinfo
  {volume} {46}},\ \href {https://doi.org/10.1140/epjh/s13129-021-00029-2}
  {10.1140/epjh/s13129-021-00029-2} (\bibinfo {year} {2021})\BibitemShut
  {NoStop}%
\bibitem [{\citenamefont {Bennett}(1987)}]{Bennett1987}%
  \BibitemOpen
  \bibfield  {author} {\bibinfo {author} {\bibfnamefont {C.~H.}\ \bibnamefont
  {Bennett}},\ }\bibfield  {title} {\bibinfo {title} {Demons, engines and the
  second law},\ }\href {https://doi.org/10.1038/scientificamerican1187-108}
  {\bibfield  {journal} {\bibinfo  {journal} {Scientific American}\ }\textbf
  {\bibinfo {volume} {257}},\ \bibinfo {pages} {108} (\bibinfo {year}
  {1987})}\BibitemShut {NoStop}%
\bibitem [{\citenamefont {Liboff}(1997)}]{Liboff1997}%
  \BibitemOpen
  \bibfield  {author} {\bibinfo {author} {\bibfnamefont {R.~L.}\ \bibnamefont
  {Liboff}},\ }\bibfield  {title} {\bibinfo {title} {Maxwell’s demon and the
  second law of thermodynamics},\ }\href {https://doi.org/10.1007/bf02764123}
  {\bibfield  {journal} {\bibinfo  {journal} {Foundations of Physics Letters}\
  }\textbf {\bibinfo {volume} {10}},\ \bibinfo {pages} {89} (\bibinfo {year}
  {1997})}\BibitemShut {NoStop}%
\bibitem [{\citenamefont {Leff}(2002)}]{Leff2002}%
  \BibitemOpen
  \bibfield  {author} {\bibinfo {author} {\bibfnamefont {H.~S.}\ \bibnamefont
  {Leff}},\ }\bibfield  {title} {\bibinfo {title} {Maxwell’s demon and the
  second law},\ }in\ \href {https://doi.org/10.1063/1.1523837} {\emph {\bibinfo
  {booktitle} {AIP Conference Proceedings}}},\ Vol.\ \bibinfo {volume} {643}\
  (\bibinfo  {publisher} {AIP},\ \bibinfo {year} {2002})\ pp.\ \bibinfo {pages}
  {408--419}\BibitemShut {NoStop}%
\bibitem [{\citenamefont {Kieu}(2004)}]{Kieu2004}%
  \BibitemOpen
  \bibfield  {author} {\bibinfo {author} {\bibfnamefont {T.~D.}\ \bibnamefont
  {Kieu}},\ }\bibfield  {title} {\bibinfo {title} {The second law,
  {M}axwell’s demon, and work derivable from quantum heat engines},\ }\href
  {https://doi.org/10.1103/physrevlett.93.140403} {\bibfield  {journal}
  {\bibinfo  {journal} {Physical Review Letters}\ }\textbf {\bibinfo {volume}
  {93}},\ \bibinfo {pages} {140403} (\bibinfo {year} {2004})}\BibitemShut
  {NoStop}%
\bibitem [{\citenamefont {Radhakrishnamurty}(2010)}]{Radhakrishnamurty2010}%
  \BibitemOpen
  \bibfield  {author} {\bibinfo {author} {\bibfnamefont {P.}~\bibnamefont
  {Radhakrishnamurty}},\ }\bibfield  {title} {\bibinfo {title} {Maxwell’s
  demon and the second law of thermodynamics},\ }\href
  {https://doi.org/10.1007/s12045-010-0061-1} {\bibfield  {journal} {\bibinfo
  {journal} {Resonance}\ }\textbf {\bibinfo {volume} {15}},\ \bibinfo {pages}
  {548} (\bibinfo {year} {2010})}\BibitemShut {NoStop}%
\bibitem [{\citenamefont {Ciliberto}(2021)}]{Ciliberto2021}%
  \BibitemOpen
  \bibfield  {author} {\bibinfo {author} {\bibfnamefont {S.}~\bibnamefont
  {Ciliberto}},\ }\bibinfo {title} {Landauer’s bound and {M}axwell’s
  demon},\ in\ \href {https://doi.org/10.1007/978-3-030-81480-9_3} {\emph
  {\bibinfo {booktitle} {Information Theory}}}\ (\bibinfo  {publisher}
  {Springer International Publishing},\ \bibinfo {year} {2021})\ pp.\ \bibinfo
  {pages} {87--112}\BibitemShut {NoStop}%
\bibitem [{\citenamefont {Fontana}(2022)}]{Fontana2022}%
  \BibitemOpen
  \bibfield  {author} {\bibinfo {author} {\bibfnamefont {P.~W.}\ \bibnamefont
  {Fontana}},\ }\bibfield  {title} {\bibinfo {title} {Hidden dissipation and
  irreversibility in {M}axwell's demon},\ }\href
  {https://doi.org/10.3390/e24010093} {\bibfield  {journal} {\bibinfo
  {journal} {Entropy}\ }\textbf {\bibinfo {volume} {24}},\ \bibinfo {pages}
  {93} (\bibinfo {year} {2022})}\BibitemShut {NoStop}%
\bibitem [{\citenamefont {Gibbs}(1902)}]{Gibbs_1902}%
  \BibitemOpen
  \bibfield  {author} {\bibinfo {author} {\bibfnamefont {J.}~\bibnamefont
  {Gibbs}},\ }\href {https://www.gutenberg.org/files/50992/50992-pdf.pdf}
  {\emph {\bibinfo {title} {Elementary principles in statistical mechanics}}}\
  (\bibinfo  {publisher} {Charles Scribner's sons},\ \bibinfo {year}
  {1902})\BibitemShut {NoStop}%
\bibitem [{\citenamefont {Balian}(1991)}]{Balian_1991}%
  \BibitemOpen
  \bibfield  {author} {\bibinfo {author} {\bibfnamefont {R.}~\bibnamefont
  {Balian}},\ }\href {https://doi.org/10.1007/978-3-540-45475-5} {\emph
  {\bibinfo {title} {From microphysics to macrophysics}}}\ (\bibinfo
  {publisher} {Springer Berlin Heidelberg},\ \bibinfo {year}
  {1991})\BibitemShut {NoStop}%
\bibitem [{\citenamefont {Dubs}(1942)}]{Dubs_1942}%
  \BibitemOpen
  \bibfield  {author} {\bibinfo {author} {\bibfnamefont {H.}~\bibnamefont
  {Dubs}},\ }\bibfield  {title} {\bibinfo {title} {The principle of
  insufficient reason},\ }\href {https://doi.org/10.1086/286754} {\bibfield
  {journal} {\bibinfo  {journal} {Philosophy of Science}\ }\textbf {\bibinfo
  {volume} {9}},\ \bibinfo {pages} {123} (\bibinfo {year} {1942})}\BibitemShut
  {NoStop}%
\bibitem [{\citenamefont {Keynes}(1921)}]{Keynes_1921}%
  \BibitemOpen
  \bibfield  {author} {\bibinfo {author} {\bibfnamefont {J.}~\bibnamefont
  {Keynes}},\ }\href {https://www.gutenberg.org/files/32625/32625-pdf.pdf}
  {\emph {\bibinfo {title} {A treatise on probability}}}\ (\bibinfo
  {publisher} {Macmillian},\ \bibinfo {year} {1921})\BibitemShut {NoStop}%
\bibitem [{\citenamefont {Ellis}(1850)}]{Ellis_1850}%
  \BibitemOpen
  \bibfield  {author} {\bibinfo {author} {\bibfnamefont {R.}~\bibnamefont
  {Ellis}},\ }\bibfield  {title} {\bibinfo {title} {Remarks on an alleged proof
  of the {\textquotedblleft}method of least squares,{\textquotedblright}
  contained in a late number of the edinburgh review},\ }\href
  {https://doi.org/10.1080/14786445008646622} {\bibfield  {journal} {\bibinfo
  {journal} {The London, Edinburgh, and Dublin Philosophical Magazine and
  Journal of Science}\ }\textbf {\bibinfo {volume} {37}},\ \bibinfo {pages}
  {321} (\bibinfo {year} {1850})}\BibitemShut {NoStop}%
\bibitem [{\citenamefont {Lairez}(2022)}]{Lairez_2022a}%
  \BibitemOpen
  \bibfield  {author} {\bibinfo {author} {\bibfnamefont {D.}~\bibnamefont
  {Lairez}},\ }\bibfield  {title} {\bibinfo {title} {A short derivation of
  {B}oltzmann distribution and {G}ibbs entropy formula from the fundamental
  postulate},\ }\href@noop {} {\bibfield  {journal} {\bibinfo  {journal}
  {arXiv}\ } (\bibinfo {year} {2022})},\ \Eprint
  {https://arxiv.org/abs/2211.02455v3} {2211.02455v3} \BibitemShut {NoStop}%
\bibitem [{\citenamefont {Tien}\ and\ \citenamefont
  {Lienhard}(1979)}]{Tien_1979}%
  \BibitemOpen
  \bibfield  {author} {\bibinfo {author} {\bibfnamefont {C.~L.}\ \bibnamefont
  {Tien}}\ and\ \bibinfo {author} {\bibfnamefont {J.}~\bibnamefont
  {Lienhard}},\ }\href@noop {} {\emph {\bibinfo {title} {Statistical
  thermodynamics}}}\ (\bibinfo  {publisher} {Hemisphere Pub. Corp.},\ \bibinfo
  {year} {1979})\BibitemShut {NoStop}%
\bibitem [{\citenamefont {Sekerka}(2015)}]{Sekerka_2015}%
  \BibitemOpen
  \bibfield  {author} {\bibinfo {author} {\bibfnamefont {R.}~\bibnamefont
  {Sekerka}},\ }\href {https://doi.org/10.1016/c2014-0-03233-9} {\emph
  {\bibinfo {title} {Thermal Physics}}}\ (\bibinfo  {publisher} {Elsevier},\
  \bibinfo {year} {2015})\BibitemShut {NoStop}%
\bibitem [{\citenamefont {Boltzmann}(1964)}]{Boltzmann_Lectures}%
  \BibitemOpen
  \bibfield  {author} {\bibinfo {author} {\bibfnamefont {L.}~\bibnamefont
  {Boltzmann}},\ }\href@noop {} {\emph {\bibinfo {title} {Lectures on gas
  theory}}}\ (\bibinfo  {publisher} {Dover ed., New York, NY, USA},\ \bibinfo
  {year} {1964})\BibitemShut {NoStop}%
\bibitem [{\citenamefont {Maxwell}(1867)}]{Maxwell_1867}%
  \BibitemOpen
  \bibfield  {author} {\bibinfo {author} {\bibfnamefont {J.~C.}\ \bibnamefont
  {Maxwell}},\ }\bibfield  {title} {\bibinfo {title} {On the dynamical theory
  of gases},\ }\href {https://doi.org/10.1098/rstl.1867.0004} {\bibfield
  {journal} {\bibinfo  {journal} {Philosophical Transactions of the Royal
  Society of London}\ }\textbf {\bibinfo {volume} {157}},\ \bibinfo {pages}
  {49} (\bibinfo {year} {1867})}\BibitemShut {NoStop}%
\bibitem [{\citenamefont {Cercignani}(1988)}]{Cercignani_1988}%
  \BibitemOpen
  \bibfield  {author} {\bibinfo {author} {\bibfnamefont {C.}~\bibnamefont
  {Cercignani}},\ }\href {https://doi.org/10.1007/978-1-4612-1039-9} {\emph
  {\bibinfo {title} {The {B}oltzmann Equation and Its Applications}}}\
  (\bibinfo  {publisher} {Springer New York},\ \bibinfo {year}
  {1988})\BibitemShut {NoStop}%
\bibitem [{\citenamefont {Villani}(2008)}]{Villani_2008}%
  \BibitemOpen
  \bibfield  {author} {\bibinfo {author} {\bibfnamefont {C.}~\bibnamefont
  {Villani}},\ }\bibfield  {title} {\bibinfo {title} {H-theorem and beyond:
  {B}oltzmann's entropy in today's mathematics},\ }in\ \href
  {https://doi.org/10.4171/057-1/9} {\emph {\bibinfo {booktitle} {{B}oltzmann's
  Legacy}}}\ (\bibinfo  {publisher} {{EMS} Press},\ \bibinfo {year} {2008})\
  pp.\ \bibinfo {pages} {129--143}\BibitemShut {NoStop}%
\bibitem [{\citenamefont {Weaver}(2021)}]{Weaver_2021}%
  \BibitemOpen
  \bibfield  {author} {\bibinfo {author} {\bibfnamefont {C.}~\bibnamefont
  {Weaver}},\ }\bibfield  {title} {\bibinfo {title} {In praise of {C}lausius
  entropy: Reassessing the foundations of {B}oltzmannian statistical
  mechanics},\ }\bibfield  {journal} {\bibinfo  {journal} {Foundations of
  Physics}\ }\textbf {\bibinfo {volume} {51}},\ \href
  {https://doi.org/10.1007/s10701-021-00437-w} {10.1007/s10701-021-00437-w}
  (\bibinfo {year} {2021})\BibitemShut {NoStop}%
\bibitem [{\citenamefont {Weaver}(2022)}]{Weaver_2022}%
  \BibitemOpen
  \bibfield  {author} {\bibinfo {author} {\bibfnamefont {C.}~\bibnamefont
  {Weaver}},\ }\bibfield  {title} {\bibinfo {title} {{P}oincar{\'{e}},
  {P}oincar{\'{e}} recurrence and the {H}-theorem: A continued reassessment of
  {B}oltzmannian statistical mechanics},\ }\bibfield  {journal} {\bibinfo
  {journal} {International Journal of Modern Physics B}\ }\textbf {\bibinfo
  {volume} {36}},\ \href {https://doi.org/10.1142/s0217979222300055}
  {10.1142/s0217979222300055} (\bibinfo {year} {2022})\BibitemShut {NoStop}%
\bibitem [{\citenamefont {Villani}(2002)}]{Villani_2002}%
  \BibitemOpen
  \bibfield  {author} {\bibinfo {author} {\bibfnamefont {C.}~\bibnamefont
  {Villani}},\ }\bibfield  {title} {\bibinfo {title} {Chap. 2 - a review of
  mathematical topics in collisional kinetic theory},\ }in\ \href
  {https://doi.org/10.1016/S1874-5792(02)80004-0} {\emph {\bibinfo {booktitle}
  {Handbook of Mathematical Fluid Dynamics}}},\ \bibinfo {series} {Handbook of
  Mathematical Fluid Dynamics}, Vol.~\bibinfo {volume} {1},\ \bibinfo {editor}
  {edited by\ \bibinfo {editor} {\bibfnamefont {S.}~\bibnamefont
  {Friedlander}}\ and\ \bibinfo {editor} {\bibfnamefont {D.}~\bibnamefont
  {Serre}}}\ (\bibinfo  {publisher} {North-Holland},\ \bibinfo {year} {2002})\
  pp.\ \bibinfo {pages} {71--74}\BibitemShut {NoStop}%
\bibitem [{\citenamefont {Uffink}(2007)}]{Uffink_2006}%
  \BibitemOpen
  \bibfield  {author} {\bibinfo {author} {\bibfnamefont {J.}~\bibnamefont
  {Uffink}},\ }\bibfield  {title} {\bibinfo {title} {Compendium of the
  foundations of classical statistical physics},\ }in\ \href
  {https://doi.org/https://doi.org/10.1016/B978-044451560-5/50012-9} {\emph
  {\bibinfo {booktitle} {Philosophy of Physics}}},\ \bibinfo {series and
  number} {Handbook of the Philosophy of Science},\ \bibinfo {editor} {edited
  by\ \bibinfo {editor} {\bibfnamefont {J.}~\bibnamefont {Butterfield}}\ and\
  \bibinfo {editor} {\bibfnamefont {J.}~\bibnamefont {Earman}}}\ (\bibinfo
  {publisher} {North-Holland},\ \bibinfo {address} {Amsterdam},\ \bibinfo
  {year} {2007})\ pp.\ \bibinfo {pages} {923--1074}\BibitemShut {NoStop}%
\bibitem [{\citenamefont {Callen}(1985)}]{Callen_1985}%
  \BibitemOpen
  \bibfield  {author} {\bibinfo {author} {\bibfnamefont {H.~B.}\ \bibnamefont
  {Callen}},\ }\href@noop {} {\emph {\bibinfo {title} {Thermodynamics and an
  introduction to thermostatistics}}},\ \bibinfo {edition} {2nd}\ ed.\
  (\bibinfo  {publisher} {J. Wiley \& sons},\ \bibinfo {year}
  {1985})\BibitemShut {NoStop}%
\bibitem [{\citenamefont {Villani}(2012)}]{Villani_2012}%
  \BibitemOpen
  \bibfield  {author} {\bibinfo {author} {\bibfnamefont {C.}~\bibnamefont
  {Villani}},\ }\bibfield  {title} {\bibinfo {title} {({I}r)reversibility and
  entropy},\ }in\ \href {https://doi.org/10.1007/978-3-0348-0359-5_2} {\emph
  {\bibinfo {booktitle} {Time}}}\ (\bibinfo  {publisher} {Springer Basel},\
  \bibinfo {year} {2012})\ pp.\ \bibinfo {pages} {19--79}\BibitemShut {NoStop}%
\bibitem [{\citenamefont {Ray}(1984)}]{Ray_1984}%
  \BibitemOpen
  \bibfield  {author} {\bibinfo {author} {\bibfnamefont {J.~R.}\ \bibnamefont
  {Ray}},\ }\bibfield  {title} {\bibinfo {title} {Correct {B}oltzmann
  counting},\ }\href {https://doi.org/10.1088/0143-0807/5/4/006} {\bibfield
  {journal} {\bibinfo  {journal} {European Journal of Physics}\ }\textbf
  {\bibinfo {volume} {5}},\ \bibinfo {pages} {219} (\bibinfo {year}
  {1984})}\BibitemShut {NoStop}%
\bibitem [{\citenamefont {Huang}(1991)}]{Huang_1987}%
  \BibitemOpen
  \bibfield  {author} {\bibinfo {author} {\bibfnamefont {K.}~\bibnamefont
  {Huang}},\ }\href
  {https://www.wiley.com/en-us/Statistical+Mechanics%2C+2nd+Edition-p-9780471815181}
  {\emph {\bibinfo {title} {{S}tatistical {M}echanics}}},\ \bibinfo {edition}
  {2nd}\ ed.\ (\bibinfo  {publisher} {J. Wiley \& sons},\ \bibinfo {year}
  {1991})\BibitemShut {NoStop}%
\bibitem [{\citenamefont {Nagle}(2004)}]{Nagle_2004}%
  \BibitemOpen
  \bibfield  {author} {\bibinfo {author} {\bibfnamefont {J.~F.}\ \bibnamefont
  {Nagle}},\ }\bibfield  {title} {\bibinfo {title} {Regarding the entropy of
  distinguishable particles},\ }\href
  {https://doi.org/10.1007/s10955-004-5715-5} {\bibfield  {journal} {\bibinfo
  {journal} {Journal of Statistical Physics}\ }\textbf {\bibinfo {volume}
  {117}},\ \bibinfo {pages} {1047} (\bibinfo {year} {2004})}\BibitemShut
  {NoStop}%
\bibitem [{\citenamefont {Cheng}(2009)}]{Cheng_2009}%
  \BibitemOpen
  \bibfield  {author} {\bibinfo {author} {\bibfnamefont {C.-H.}\ \bibnamefont
  {Cheng}},\ }\bibfield  {title} {\bibinfo {title} {Thermodynamics of the
  system of distinguishable particles},\ }\href
  {https://doi.org/10.3390/e11030326} {\bibfield  {journal} {\bibinfo
  {journal} {Entropy}\ }\textbf {\bibinfo {volume} {11}},\ \bibinfo {pages}
  {326} (\bibinfo {year} {2009})}\BibitemShut {NoStop}%
\bibitem [{\citenamefont {Versteegh}\ and\ \citenamefont
  {Dieks}(2011)}]{Versteegh2011}%
  \BibitemOpen
  \bibfield  {author} {\bibinfo {author} {\bibfnamefont {M.~A.~M.}\
  \bibnamefont {Versteegh}}\ and\ \bibinfo {author} {\bibfnamefont
  {D.}~\bibnamefont {Dieks}},\ }\bibfield  {title} {\bibinfo {title} {The
  {G}ibbs paradox and the distinguishability of identical particles},\ }\href
  {https://doi.org/10.1119/1.3584179} {\bibfield  {journal} {\bibinfo
  {journal} {American Journal of Physics}\ }\textbf {\bibinfo {volume} {79}},\
  \bibinfo {pages} {741} (\bibinfo {year} {2011})}\BibitemShut {NoStop}%
\bibitem [{\citenamefont {Frenkel}(2014)}]{Frenkel_2014}%
  \BibitemOpen
  \bibfield  {author} {\bibinfo {author} {\bibfnamefont {D.}~\bibnamefont
  {Frenkel}},\ }\bibfield  {title} {\bibinfo {title} {Why colloidal systems can
  be described by statistical mechanics: some not very original comments on the
  {G}ibbs paradox},\ }\href {https://doi.org/10.1080/00268976.2014.904051}
  {\bibfield  {journal} {\bibinfo  {journal} {Molecular Physics}\ }\textbf
  {\bibinfo {volume} {112}},\ \bibinfo {pages} {2325} (\bibinfo {year}
  {2014})}\BibitemShut {NoStop}%
\bibitem [{\citenamefont {Dieks}(2018)}]{Dieks_2018}%
  \BibitemOpen
  \bibfield  {author} {\bibinfo {author} {\bibfnamefont {D.}~\bibnamefont
  {Dieks}},\ }\bibfield  {title} {\bibinfo {title} {The {G}ibbs paradox and
  particle individuality},\ }\href {https://doi.org/10.3390/e20060466}
  {\bibfield  {journal} {\bibinfo  {journal} {Entropy}\ }\textbf {\bibinfo
  {volume} {20}},\ \bibinfo {pages} {466} (\bibinfo {year} {2018})}\BibitemShut
  {NoStop}%
\bibitem [{\citenamefont {Saunders}(2018)}]{Saunders_2018}%
  \BibitemOpen
  \bibfield  {author} {\bibinfo {author} {\bibfnamefont {S.}~\bibnamefont
  {Saunders}},\ }\bibfield  {title} {\bibinfo {title} {The {G}ibbs paradox},\
  }\href {https://doi.org/10.3390/e20080552} {\bibfield  {journal} {\bibinfo
  {journal} {Entropy}\ }\textbf {\bibinfo {volume} {20}},\ \bibinfo {pages}
  {552} (\bibinfo {year} {2018})}\BibitemShut {NoStop}%
\bibitem [{\citenamefont {van Lith}(2018)}]{van_Lith_2018}%
  \BibitemOpen
  \bibfield  {author} {\bibinfo {author} {\bibfnamefont {J.}~\bibnamefont {van
  Lith}},\ }\bibfield  {title} {\bibinfo {title} {The {G}ibbs paradox: lessons
  from thermodynamics},\ }\href {https://doi.org/10.3390/e20050328} {\bibfield
  {journal} {\bibinfo  {journal} {Entropy}\ }\textbf {\bibinfo {volume} {20}},\
  \bibinfo {pages} {328} (\bibinfo {year} {2018})}\BibitemShut {NoStop}%
\bibitem [{\citenamefont {Lairez}(2023{\natexlab{a}})}]{Lairez_Stirling}%
  \BibitemOpen
  \bibfield  {author} {\bibinfo {author} {\bibfnamefont {D.}~\bibnamefont
  {Lairez}},\ }\bibfield  {title} {\bibinfo {title} {Plea for the use of the
  exact {S}tirling formula in statistical mechanics},\ }\bibfield  {journal}
  {\bibinfo  {journal} {SciPost Physics Lecture Notes}\ }\href
  {https://doi.org/10.21468/scipostphyslectnotes.76}
  {10.21468/scipostphyslectnotes.76} (\bibinfo {year}
  {2023}{\natexlab{a}})\BibitemShut {NoStop}%
\bibitem [{\citenamefont {Kobayashi}\ \emph {et~al.}(2025)\citenamefont
  {Kobayashi}, \citenamefont {Park}, \citenamefont {Miné-Hattab},\ and\
  \citenamefont {Fernández}}]{Kobayashi2025}%
  \BibitemOpen
  \bibfield  {author} {\bibinfo {author} {\bibfnamefont {A.}~\bibnamefont
  {Kobayashi}}, \bibinfo {author} {\bibfnamefont {J.}~\bibnamefont {Park}},
  \bibinfo {author} {\bibfnamefont {J.}~\bibnamefont {Miné-Hattab}},\ and\
  \bibinfo {author} {\bibfnamefont {F.~G.}\ \bibnamefont {Fernández}},\ }\href
  {https://doi.org/10.1101/2025.04.03.647105} {\bibinfo {title} {A guide for
  single particle tracking: from sample preparation and image acquisition to
  the analysis of individual trajectories}} (\bibinfo {year}
  {2025})\BibitemShut {NoStop}%
\bibitem [{\citenamefont {Crocker}\ and\ \citenamefont
  {Hoffman}(2007)}]{Crocker2007}%
  \BibitemOpen
  \bibfield  {author} {\bibinfo {author} {\bibfnamefont {J.~C.}\ \bibnamefont
  {Crocker}}\ and\ \bibinfo {author} {\bibfnamefont {B.~D.}\ \bibnamefont
  {Hoffman}},\ }\bibinfo {title} {Multiple‐particle tracking and two‐point
  microrheology in cells},\ in\ \href
  {https://doi.org/10.1016/s0091-679x(07)83007-x} {\emph {\bibinfo {booktitle}
  {Cell Mechanics}}}\ (\bibinfo  {publisher} {Elsevier},\ \bibinfo {year}
  {2007})\ pp.\ \bibinfo {pages} {141--178}\BibitemShut {NoStop}%
\bibitem [{\citenamefont {Pinto}\ \emph {et~al.}(2021)\citenamefont {Pinto},
  \citenamefont {Vickers}, \citenamefont {Sharifi},\ and\ \citenamefont
  {Andersson}}]{Pinto2021}%
  \BibitemOpen
  \bibfield  {author} {\bibinfo {author} {\bibfnamefont {S.~C.}\ \bibnamefont
  {Pinto}}, \bibinfo {author} {\bibfnamefont {N.~A.}\ \bibnamefont {Vickers}},
  \bibinfo {author} {\bibfnamefont {F.}~\bibnamefont {Sharifi}},\ and\ \bibinfo
  {author} {\bibfnamefont {S.~B.}\ \bibnamefont {Andersson}},\ }\bibfield
  {title} {\bibinfo {title} {Tracking multiple diffusing particles using
  information optimal control},\ }in\ \href
  {https://doi.org/10.23919/acc50511.2021.9482619} {\emph {\bibinfo {booktitle}
  {2021 American Control Conference (ACC)}}}\ (\bibinfo  {publisher} {IEEE},\
  \bibinfo {year} {2021})\ pp.\ \bibinfo {pages} {4033--4038}\BibitemShut
  {NoStop}%
\bibitem [{\citenamefont {Xu}\ \emph {et~al.}(2024)\citenamefont {Xu},
  \citenamefont {Wei},\ and\ \citenamefont {Sang}}]{Xu2024}%
  \BibitemOpen
  \bibfield  {author} {\bibinfo {author} {\bibfnamefont {X.}~\bibnamefont
  {Xu}}, \bibinfo {author} {\bibfnamefont {J.}~\bibnamefont {Wei}},\ and\
  \bibinfo {author} {\bibfnamefont {S.}~\bibnamefont {Sang}},\ }\bibfield
  {title} {\bibinfo {title} {Deep learning-based multiple particle tracking in
  complex system},\ }\bibfield  {journal} {\bibinfo  {journal} {AIP Advances}\
  }\textbf {\bibinfo {volume} {14}},\ \href {https://doi.org/10.1063/5.0186670}
  {10.1063/5.0186670} (\bibinfo {year} {2024})\BibitemShut {NoStop}%
\bibitem [{\citenamefont {Peters}(2013)}]{Peters_2013}%
  \BibitemOpen
  \bibfield  {author} {\bibinfo {author} {\bibfnamefont {H.}~\bibnamefont
  {Peters}},\ }\bibfield  {title} {\bibinfo {title} {Demonstration and
  resolution of the {G}ibbs paradox of the first kind},\ }\href
  {https://doi.org/10.1088/0143-0807/35/1/015023} {\bibfield  {journal}
  {\bibinfo  {journal} {European Journal of Physics}\ }\textbf {\bibinfo
  {volume} {35}},\ \bibinfo {pages} {015023} (\bibinfo {year}
  {2013})}\BibitemShut {NoStop}%
\bibitem [{\citenamefont {Shannon}(1948)}]{Shannon_1948}%
  \BibitemOpen
  \bibfield  {author} {\bibinfo {author} {\bibfnamefont {C.~E.}\ \bibnamefont
  {Shannon}},\ }\bibfield  {title} {\bibinfo {title} {A mathematical theory of
  communication},\ }\href {https://doi.org/10.1002/j.1538-7305.1948.tb01338.x}
  {\bibfield  {journal} {\bibinfo  {journal} {The Bell System Technical
  Journal}\ }\textbf {\bibinfo {volume} {27}},\ \bibinfo {pages} {379}
  (\bibinfo {year} {1948})}\BibitemShut {NoStop}%
\bibitem [{\citenamefont {Hartley}(1928)}]{Hartley_1928}%
  \BibitemOpen
  \bibfield  {author} {\bibinfo {author} {\bibfnamefont {R.~V.~L.}\
  \bibnamefont {Hartley}},\ }\bibfield  {title} {\bibinfo {title} {Transmission
  of information},\ }\href {https://doi.org/10.1002/j.1538-7305.1928.tb01236.x}
  {\bibfield  {journal} {\bibinfo  {journal} {Bell System Technical Journal}\
  }\textbf {\bibinfo {volume} {7}},\ \bibinfo {pages} {535} (\bibinfo {year}
  {1928})}\BibitemShut {NoStop}%
\bibitem [{\citenamefont {Jaynes}(1957)}]{Jaynes_1957}%
  \BibitemOpen
  \bibfield  {author} {\bibinfo {author} {\bibfnamefont {E.~T.}\ \bibnamefont
  {Jaynes}},\ }\bibfield  {title} {\bibinfo {title} {Information theory and
  statistical mechanics},\ }\href {https://doi.org/10.1103/PhysRev.106.620}
  {\bibfield  {journal} {\bibinfo  {journal} {Phys. Rev.}\ }\textbf {\bibinfo
  {volume} {106}},\ \bibinfo {pages} {620} (\bibinfo {year}
  {1957})}\BibitemShut {NoStop}%
\bibitem [{\citenamefont {Shore}\ and\ \citenamefont
  {Johnson}(1980)}]{Shore_1980}%
  \BibitemOpen
  \bibfield  {author} {\bibinfo {author} {\bibfnamefont {J.}~\bibnamefont
  {Shore}}\ and\ \bibinfo {author} {\bibfnamefont {R.}~\bibnamefont
  {Johnson}},\ }\bibfield  {title} {\bibinfo {title} {Axiomatic derivation of
  the principle of maximum entropy and the principle of minimum
  cross-entropy},\ }\href {https://doi.org/10.1109/tit.1980.1056144} {\bibfield
   {journal} {\bibinfo  {journal} {{IEEE} Transactions on Information Theory}\
  }\textbf {\bibinfo {volume} {26}},\ \bibinfo {pages} {26} (\bibinfo {year}
  {1980})}\BibitemShut {NoStop}%
\bibitem [{\citenamefont {Jaynes}(1973)}]{Jaynes_1973}%
  \BibitemOpen
  \bibfield  {author} {\bibinfo {author} {\bibfnamefont {E.~T.}\ \bibnamefont
  {Jaynes}},\ }\bibfield  {title} {\bibinfo {title} {The well-posed problem},\
  }\href {https://doi.org/10.1007/bf00709116} {\bibfield  {journal} {\bibinfo
  {journal} {Foundations of Physics}\ }\textbf {\bibinfo {volume} {3}},\
  \bibinfo {pages} {477} (\bibinfo {year} {1973})}\BibitemShut {NoStop}%
\bibitem [{\citenamefont {Brillouin}(1953)}]{Brillouin1953}%
  \BibitemOpen
  \bibfield  {author} {\bibinfo {author} {\bibfnamefont {L.}~\bibnamefont
  {Brillouin}},\ }\bibfield  {title} {\bibinfo {title} {The negentropy
  principle of information},\ }\href {https://doi.org/10.1063/1.1721463}
  {\bibfield  {journal} {\bibinfo  {journal} {Journal of Applied Physics}\
  }\textbf {\bibinfo {volume} {24}},\ \bibinfo {pages} {1152} (\bibinfo {year}
  {1953})}\BibitemShut {NoStop}%
\bibitem [{\citenamefont {Brillouin}(1956)}]{Brillouin_1956_book}%
  \BibitemOpen
  \bibfield  {author} {\bibinfo {author} {\bibfnamefont {L.}~\bibnamefont
  {Brillouin}},\ }\href@noop {} {\emph {\bibinfo {title} {Science and
  Information Theory}}}\ (\bibinfo  {publisher} {Dover Publications},\ \bibinfo
  {address} {Mineola, N.Y.},\ \bibinfo {year} {1956})\BibitemShut {NoStop}%
\bibitem [{\citenamefont {Ciliberto}(2018)}]{Ciliberto_2018a}%
  \BibitemOpen
  \bibfield  {author} {\bibinfo {author} {\bibfnamefont {S.}~\bibnamefont
  {Ciliberto}},\ }\bibfield  {title} {\bibinfo {title} {Landauer's bound and
  maxwell's demon},\ }\href@noop {} {\bibfield  {journal} {\bibinfo  {journal}
  {S\'eminaire Poincar\'e}\ }\textbf {\bibinfo {volume} {L'Information,
  XXIII}},\ \bibinfo {pages} {79} (\bibinfo {year} {2018})}\BibitemShut
  {NoStop}%
\bibitem [{\citenamefont {Penrose}(1979)}]{Penrose_1979}%
  \BibitemOpen
  \bibfield  {author} {\bibinfo {author} {\bibfnamefont {O.}~\bibnamefont
  {Penrose}},\ }\bibfield  {title} {\bibinfo {title} {Foundations of
  statistical mechanics},\ }\href {https://doi.org/10.1088/0034-4885/42/12/002}
  {\bibfield  {journal} {\bibinfo  {journal} {Reports on Progress in Physics}\
  }\textbf {\bibinfo {volume} {42}},\ \bibinfo {pages} {1937} (\bibinfo {year}
  {1979})}\BibitemShut {NoStop}%
\bibitem [{\citenamefont {Denbigh}\ and\ \citenamefont
  {Denbigh}(1985)}]{Denbigh_1985}%
  \BibitemOpen
  \bibfield  {author} {\bibinfo {author} {\bibfnamefont {K.}~\bibnamefont
  {Denbigh}}\ and\ \bibinfo {author} {\bibfnamefont {J.}~\bibnamefont
  {Denbigh}},\ }\href {https://doi.org/10.1119/1.14692} {\emph {\bibinfo
  {title} {Entropy in Relation to Incomplete Knowledge}}}\ (\bibinfo
  {publisher} {Cambridge University Press},\ \bibinfo {year}
  {1985})\BibitemShut {NoStop}%
\bibitem [{\citenamefont {Lavis}(2007)}]{Lavis_2015}%
  \BibitemOpen
  \bibfield  {author} {\bibinfo {author} {\bibfnamefont {D.~A.}\ \bibnamefont
  {Lavis}},\ }\bibinfo {title} {Frontiers in fundamental physics},\ in\
  \href@noop {} {\emph {\bibinfo {booktitle} {Frontiers in Fundamental
  Physics}}},\ Vol.\ \bibinfo {volume} {vol 3},\ \bibinfo {editor} {edited by\
  \bibinfo {editor} {\bibfnamefont {B.~G.}\ \bibnamefont {Sidarth}}}\ (\bibinfo
   {publisher} {Universities Press, India,},\ \bibinfo {year} {2007})\ Chap.\
  \bibinfo {chapter} {Equilibrium and (Ir)reversibility in Classical
  Statistical Mechanics}\BibitemShut {NoStop}%
\bibitem [{\citenamefont {Landauer}(1991)}]{Landauer_1991}%
  \BibitemOpen
  \bibfield  {author} {\bibinfo {author} {\bibfnamefont {R.}~\bibnamefont
  {Landauer}},\ }\bibfield  {title} {\bibinfo {title} {Information is
  physical},\ }\href {https://doi.org/10.1063/1.881299} {\bibfield  {journal}
  {\bibinfo  {journal} {Physics Today}\ }\textbf {\bibinfo {volume} {44}},\
  \bibinfo {pages} {23} (\bibinfo {year} {1991})}\BibitemShut {NoStop}%
\bibitem [{\citenamefont {Landauer}(1961)}]{Landauer_1961}%
  \BibitemOpen
  \bibfield  {author} {\bibinfo {author} {\bibfnamefont {R.}~\bibnamefont
  {Landauer}},\ }\bibfield  {title} {\bibinfo {title} {Irreversibility and heat
  generation in the computing process},\ }\href
  {https://doi.org/10.1147/rd.53.0183} {\bibfield  {journal} {\bibinfo
  {journal} {{IBM} Journal of Research and Development}\ }\textbf {\bibinfo
  {volume} {5}},\ \bibinfo {pages} {183} (\bibinfo {year} {1961})}\BibitemShut
  {NoStop}%
\bibitem [{\citenamefont {Landauer}(1996)}]{Landauer1996}%
  \BibitemOpen
  \bibfield  {author} {\bibinfo {author} {\bibfnamefont {R.}~\bibnamefont
  {Landauer}},\ }\bibfield  {title} {\bibinfo {title} {The physical nature of
  information},\ }\href {https://doi.org/10.1016/0375-9601(96)00453-7}
  {\bibfield  {journal} {\bibinfo  {journal} {Physics Letters A}\ }\textbf
  {\bibinfo {volume} {217}},\ \bibinfo {pages} {188} (\bibinfo {year}
  {1996})}\BibitemShut {NoStop}%
\bibitem [{\citenamefont {Bennett}(1982)}]{Bennett_1982}%
  \BibitemOpen
  \bibfield  {author} {\bibinfo {author} {\bibfnamefont {C.~H.}\ \bibnamefont
  {Bennett}},\ }\bibfield  {title} {\bibinfo {title} {The thermodynamics of
  computation - a review},\ }\href {https://doi.org/10.1007/bf02084158}
  {\bibfield  {journal} {\bibinfo  {journal} {International Journal of
  Theoretical Physics}\ }\textbf {\bibinfo {volume} {21}},\ \bibinfo {pages}
  {905} (\bibinfo {year} {1982})}\BibitemShut {NoStop}%
\bibitem [{\citenamefont {Bennett}(2003)}]{Bennett_2003}%
  \BibitemOpen
  \bibfield  {author} {\bibinfo {author} {\bibfnamefont {C.~H.}\ \bibnamefont
  {Bennett}},\ }\bibfield  {title} {\bibinfo {title} {Notes on
  {L}andauer{\textquotesingle}s principle, reversible computation, and
  {M}axwell{\textquotesingle}s demon},\ }\href
  {https://doi.org/10.1016/s1355-2198(03)00039-x} {\bibfield  {journal}
  {\bibinfo  {journal} {Studies in History and Philosophy of Science Part B:
  Studies in History and Philosophy of Modern Physics}\ }\textbf {\bibinfo
  {volume} {34}},\ \bibinfo {pages} {501} (\bibinfo {year} {2003})}\BibitemShut
  {NoStop}%
\bibitem [{\citenamefont {Herrera}(2020)}]{Herrera2020}%
  \BibitemOpen
  \bibfield  {author} {\bibinfo {author} {\bibfnamefont {L.}~\bibnamefont
  {Herrera}},\ }\bibfield  {title} {\bibinfo {title} {Landauer principle and
  general relativity},\ }\href {https://doi.org/10.3390/e22030340} {\bibfield
  {journal} {\bibinfo  {journal} {Entropy}\ }\textbf {\bibinfo {volume} {22}},\
  \bibinfo {pages} {340} (\bibinfo {year} {2020})}\BibitemShut {NoStop}%
\bibitem [{\citenamefont {Lutz}\ and\ \citenamefont
  {Ciliberto}(2015)}]{Lutz_2015}%
  \BibitemOpen
  \bibfield  {author} {\bibinfo {author} {\bibfnamefont {E.}~\bibnamefont
  {Lutz}}\ and\ \bibinfo {author} {\bibfnamefont {S.}~\bibnamefont
  {Ciliberto}},\ }\bibfield  {title} {\bibinfo {title} {Information: from
  {M}axwell's demon to {L}andauer's eraser},\ }\href
  {https://doi.org/10.1063/pt.3.2912} {\bibfield  {journal} {\bibinfo
  {journal} {Physics Today}\ }\textbf {\bibinfo {volume} {68}},\ \bibinfo
  {pages} {30} (\bibinfo {year} {2015})}\BibitemShut {NoStop}%
\bibitem [{\citenamefont {Bormashenko}(2019)}]{Bormashenko_2019b}%
  \BibitemOpen
  \bibfield  {author} {\bibinfo {author} {\bibfnamefont {E.}~\bibnamefont
  {Bormashenko}},\ }\bibfield  {title} {\bibinfo {title} {The {L}andauer
  principle: Re-formulation of the second thermodynamics law or a step to great
  unification?},\ }\href {https://doi.org/10.3390/e21100918} {\bibfield
  {journal} {\bibinfo  {journal} {Entropy}\ }\textbf {\bibinfo {volume} {21}},\
  \bibinfo {pages} {918} (\bibinfo {year} {2019})}\BibitemShut {NoStop}%
\bibitem [{\citenamefont {Witkowski}\ \emph {et~al.}(2024)\citenamefont
  {Witkowski}, \citenamefont {Brown},\ and\ \citenamefont
  {Truong}}]{Witkowski2024}%
  \BibitemOpen
  \bibfield  {author} {\bibinfo {author} {\bibfnamefont {C.}~\bibnamefont
  {Witkowski}}, \bibinfo {author} {\bibfnamefont {S.}~\bibnamefont {Brown}},\
  and\ \bibinfo {author} {\bibfnamefont {K.}~\bibnamefont {Truong}},\
  }\bibfield  {title} {\bibinfo {title} {On the precise link between energy and
  information},\ }\href {https://doi.org/10.3390/e26030203} {\bibfield
  {journal} {\bibinfo  {journal} {Entropy}\ }\textbf {\bibinfo {volume} {26}},\
  \bibinfo {pages} {203} (\bibinfo {year} {2024})}\BibitemShut {NoStop}%
\bibitem [{\citenamefont {Herrera}(2014)}]{HERRERA2014}%
  \BibitemOpen
  \bibfield  {author} {\bibinfo {author} {\bibfnamefont {L.}~\bibnamefont
  {Herrera}},\ }\bibfield  {title} {\bibinfo {title} {The mass of a bit of
  information and the {B}rillouin's principle},\ }\href
  {https://doi.org/10.1142/s0219477514500023} {\bibfield  {journal} {\bibinfo
  {journal} {Fluctuation and Noise Letters}\ }\textbf {\bibinfo {volume}
  {13}},\ \bibinfo {pages} {1450002} (\bibinfo {year} {2014})}\BibitemShut
  {NoStop}%
\bibitem [{\citenamefont {Vopson}(2019)}]{Vopson_2019}%
  \BibitemOpen
  \bibfield  {author} {\bibinfo {author} {\bibfnamefont {M.~M.}\ \bibnamefont
  {Vopson}},\ }\bibfield  {title} {\bibinfo {title} {The
  mass-energy-information equivalence principle},\ }\href
  {https://doi.org/10.1063/1.5123794} {\bibfield  {journal} {\bibinfo
  {journal} {{AIP} Advances}\ }\textbf {\bibinfo {volume} {9}},\ \bibinfo
  {pages} {095206} (\bibinfo {year} {2019})}\BibitemShut {NoStop}%
\bibitem [{\citenamefont {Vopson}(2020)}]{Vopson_2020}%
  \BibitemOpen
  \bibfield  {author} {\bibinfo {author} {\bibfnamefont {M.~M.}\ \bibnamefont
  {Vopson}},\ }\bibfield  {title} {\bibinfo {title} {The information
  catastrophe},\ }\bibfield  {journal} {\bibinfo  {journal} {AIP Advances}\
  }\textbf {\bibinfo {volume} {10}},\ \href {https://doi.org/10.1063/5.0019941}
  {10.1063/5.0019941} (\bibinfo {year} {2020})\BibitemShut {NoStop}%
\bibitem [{\citenamefont {D\v{z}aferovi\'c-Ma\v{s}i\'c}(2021)}]{Masic_2021}%
  \BibitemOpen
  \bibfield  {author} {\bibinfo {author} {\bibfnamefont {E.}~\bibnamefont
  {D\v{z}aferovi\'c-Ma\v{s}i\'c}},\ }\bibfield  {title} {\bibinfo {title}
  {Missing information in the universe as a dark matter candidate based on the
  mass-energy-information equivalence principle},\ }\href
  {https://doi.org/10.1088/1742-6596/1814/1/012006} {\bibfield  {journal}
  {\bibinfo  {journal} {Journal of Physics: Conference Series}\ }\textbf
  {\bibinfo {volume} {1814}},\ \bibinfo {pages} {012006} (\bibinfo {year}
  {2021})}\BibitemShut {NoStop}%
\bibitem [{\citenamefont {Vopson}(2022)}]{Vopson_2022}%
  \BibitemOpen
  \bibfield  {author} {\bibinfo {author} {\bibfnamefont {M.~M.}\ \bibnamefont
  {Vopson}},\ }\bibfield  {title} {\bibinfo {title} {Experimental protocol for
  testing the mass-energy-information equivalence principle},\ }\href
  {https://doi.org/10.1063/5.0087175} {\bibfield  {journal} {\bibinfo
  {journal} {{AIP} Advances}\ }\textbf {\bibinfo {volume} {12}},\ \bibinfo
  {pages} {035311} (\bibinfo {year} {2022})}\BibitemShut {NoStop}%
\bibitem [{\citenamefont {Ciliberto}\ and\ \citenamefont
  {Lutz}(2018)}]{Ciliberto_2018}%
  \BibitemOpen
  \bibfield  {author} {\bibinfo {author} {\bibfnamefont {S.}~\bibnamefont
  {Ciliberto}}\ and\ \bibinfo {author} {\bibfnamefont {E.}~\bibnamefont
  {Lutz}},\ }\bibfield  {title} {\bibinfo {title} {The physics of information:
  from {M}axwell to {L}andauer},\ }in\ \href
  {https://doi.org/10.1007/978-3-319-93458-7_5} {\emph {\bibinfo {booktitle}
  {Energy Limits in Computation}}}\ (\bibinfo  {publisher} {Springer
  International Publishing},\ \bibinfo {year} {2018})\ pp.\ \bibinfo {pages}
  {155--175}\BibitemShut {NoStop}%
\bibitem [{\citenamefont {Lairez}(2024{\natexlab{b}})}]{Lairez_2024b}%
  \BibitemOpen
  \bibfield  {author} {\bibinfo {author} {\bibfnamefont {D.}~\bibnamefont
  {Lairez}},\ }\bibfield  {title} {\bibinfo {title} {The fundamental difference
  between boolean logic and thermodynamic irreversibilities, or, why
  {L}andauer’s result cannot be a physical principle},\ }\href
  {https://doi.org/10.3390/sym16121594} {\bibfield  {journal} {\bibinfo
  {journal} {Symmetry}\ }\textbf {\bibinfo {volume} {16}},\ \bibinfo {pages}
  {1594} (\bibinfo {year} {2024}{\natexlab{b}})}\BibitemShut {NoStop}%
\bibitem [{\citenamefont {Plenio}\ and\ \citenamefont
  {Vitelli}(2001)}]{Plenio_2001}%
  \BibitemOpen
  \bibfield  {author} {\bibinfo {author} {\bibfnamefont {M.~B.}\ \bibnamefont
  {Plenio}}\ and\ \bibinfo {author} {\bibfnamefont {V.}~\bibnamefont
  {Vitelli}},\ }\bibfield  {title} {\bibinfo {title} {The physics of
  forgetting: Landauer’s erasure principle and information theory},\ }\href
  {https://doi.org/10.1080/00107510010018916} {\bibfield  {journal} {\bibinfo
  {journal} {Contemporary Physics}\ }\textbf {\bibinfo {volume} {42}},\
  \bibinfo {pages} {25} (\bibinfo {year} {2001})}\BibitemShut {NoStop}%
\bibitem [{\citenamefont {Maroney}(2005)}]{Maroney2005}%
  \BibitemOpen
  \bibfield  {author} {\bibinfo {author} {\bibfnamefont {O.}~\bibnamefont
  {Maroney}},\ }\bibfield  {title} {\bibinfo {title} {The (absence of a)
  relationship between thermodynamic and logical reversibility},\ }\href
  {https://doi.org/10.1016/j.shpsb.2004.11.006} {\bibfield  {journal} {\bibinfo
   {journal} {Studies in History and Philosophy of Science Part B: Studies in
  History and Philosophy of Modern Physics}\ }\textbf {\bibinfo {volume}
  {36}},\ \bibinfo {pages} {355} (\bibinfo {year} {2005})}\BibitemShut
  {NoStop}%
\bibitem [{\citenamefont {Lairez}(2023{\natexlab{b}})}]{Lairez_2023}%
  \BibitemOpen
  \bibfield  {author} {\bibinfo {author} {\bibfnamefont {D.}~\bibnamefont
  {Lairez}},\ }\bibfield  {title} {\bibinfo {title} {Thermodynamical versus
  logical irreversibility: A concrete objection to {L}andauer's principle},\
  }\href {https://doi.org/10.3390/e25081155} {\bibfield  {journal} {\bibinfo
  {journal} {Entropy}\ }\textbf {\bibinfo {volume} {25}},\ \bibinfo {pages}
  {1155} (\bibinfo {year} {2023}{\natexlab{b}})}\BibitemShut {NoStop}%
\bibitem [{\citenamefont {Lairez}(2024{\natexlab{c}})}]{Lairez_2024a}%
  \BibitemOpen
  \bibfield  {author} {\bibinfo {author} {\bibfnamefont {D.}~\bibnamefont
  {Lairez}},\ }\bibfield  {title} {\bibinfo {title} {On the supposed mass of
  entropy and that of information},\ }\href {https://doi.org/10.3390/e26040337}
  {\bibfield  {journal} {\bibinfo  {journal} {Entropy}\ }\textbf {\bibinfo
  {volume} {26}},\ \bibinfo {pages} {337} (\bibinfo {year}
  {2024}{\natexlab{c}})}\BibitemShut {NoStop}%
\bibitem [{\citenamefont {Brillouin}(1965{\natexlab{a}})}]{Brillouin_1965a}%
  \BibitemOpen
  \bibfield  {author} {\bibinfo {author} {\bibfnamefont {L.}~\bibnamefont
  {Brillouin}},\ }\bibfield  {title} {\bibinfo {title} {The actual mass of
  potential energy, a correction to classical relativity},\ }\href
  {https://doi.org/10.1073/pnas.53.3.475} {\bibfield  {journal} {\bibinfo
  {journal} {PNAS}\ }\textbf {\bibinfo {volume} {53}},\ \bibinfo {pages} {475}
  (\bibinfo {year} {1965}{\natexlab{a}})}\BibitemShut {NoStop}%
\bibitem [{\citenamefont {Brillouin}(1965{\natexlab{b}})}]{Brillouin_1965b}%
  \BibitemOpen
  \bibfield  {author} {\bibinfo {author} {\bibfnamefont {L.}~\bibnamefont
  {Brillouin}},\ }\bibfield  {title} {\bibinfo {title} {The actual mass of
  potential energy {II}},\ }\href {https://doi.org/10.1073/pnas.53.6.1280}
  {\bibfield  {journal} {\bibinfo  {journal} {PNAS}\ }\textbf {\bibinfo
  {volume} {53}},\ \bibinfo {pages} {1280} (\bibinfo {year}
  {1965}{\natexlab{b}})}\BibitemShut {NoStop}%
\bibitem [{\citenamefont {Rosenblum}\ and\ \citenamefont
  {Kuttner}(2002)}]{Rosenblum2002}%
  \BibitemOpen
  \bibfield  {author} {\bibinfo {author} {\bibfnamefont {B.}~\bibnamefont
  {Rosenblum}}\ and\ \bibinfo {author} {\bibfnamefont {F.}~\bibnamefont
  {Kuttner}},\ }\href {https://doi.org/10.1023/a:1019723420678} {\bibfield
  {journal} {\bibinfo  {journal} {Foundations of Physics}\ }\textbf {\bibinfo
  {volume} {32}},\ \bibinfo {pages} {1273} (\bibinfo {year}
  {2002})}\BibitemShut {NoStop}%
\bibitem [{\citenamefont {Feynman}(1963)}]{Feynman_1963_relativity}%
  \BibitemOpen
  \bibfield  {author} {\bibinfo {author} {\bibfnamefont {R.}~\bibnamefont
  {Feynman}},\ }\href {https://www.feynmanlectures.caltech.edu/I_15.html}
  {\bibinfo {title} {The {F}eynman lectures on physics vol. {I}, chap. 15: The
  special theory of relativity}} (\bibinfo {year} {1963})\BibitemShut {NoStop}%
\bibitem [{\citenamefont {Bini}(2014)}]{Bini2014}%
  \BibitemOpen
  \bibfield  {author} {\bibinfo {author} {\bibfnamefont {D.}~\bibnamefont
  {Bini}},\ }\bibinfo {title} {Observers, observables and measurements in
  general relativity},\ in\ \href {https://doi.org/10.1007/978-3-319-06349-2_3}
  {\emph {\bibinfo {booktitle} {General Relativity, Cosmology and
  Astrophysics}}}\ (\bibinfo  {publisher} {Springer International Publishing},\
  \bibinfo {year} {2014})\ pp.\ \bibinfo {pages} {67--90}\BibitemShut {NoStop}%
\end{thebibliography}%
	
\end{document}